\documentclass[preprint, 12pt]{elsarticle}
\usepackage[T1]{fontenc}

\usepackage{amssymb}
\usepackage{amsmath}

\journal{Physica D: Nonlinear Phenomena}

\usepackage{xcolor}
\usepackage[unicode]{hyperref}
\usepackage{graphicx}

\def\matrix2#1{\left(\begin{array}{cc}#1\end{array}\right)}

\begin{document}

\begin{frontmatter}



\title{Discrete integrable equations with three independent variables}


\author{Ismagil T. Habibullin} 
\author{Aigul R. Khakimova}
\affiliation{organization={Ufa Institute of Mathematics, Russian Academy 
of Sciences},
            addressline={\\112 Chernyshevsky Street}, 
            city={Ufa},
            postcode={450008}, 
            country={Russia}}

\begin{abstract}
In this paper, we study nonlinear integrable equations with three independent variables of the following types: Toda-type lattices, semi-discrete lattices, and fully discrete Hirota-Miwa type models. It is shown that integrable equations of all three types admit reductions in the form of Darboux-integrable hyperbolic systems.
It is important that the transition from one class to another is carried out by means of discretization (continualization) of the above-mentioned reductions with preservation of characteristic integrals. In other words, at the level of reductions, one can establish some correspondence between the classes of 3D models under consideration. In the context of this correspondence, the authors managed to conduct a comparative analysis of the well-known list of integrable Hirota-Miwa type equations, containing 13 equations. It was established that some equations from this list are related by point changes of variables. As a result, the final list of known integrable Hirota-Miwa type equations was reduced to seven. One equation was obtained by discretizing the list of semi-discrete Toda-type equations using characteristic integrals in this paper, probably it is new. For all seven models, associated linear systems (Lax pairs) are given.

\end{abstract}


\end{frontmatter}

\allowdisplaybreaks


\section{Introduction}

The aim of this paper is to establish the relationship between the following three classes of discrete integrable equations in 3D. The first of these consists of equations of the two-dimensional Toda-type lattice:
\begin{equation}  \label{todatype}
u_{n,xy} = f(u_{n+1},u_{n},u_{n-1}, u_{n,x},u_{n,y}),
\end{equation}
where the unknown function $u_n$ depends on one discrete variable $n$ and two continuous variables $x$ and $y$. Here $u_{n,x}=\frac{\partial}{\partial x}u_n$, $u_{n,y}=\frac{\partial}{\partial y}u_n$, $u_{n,xy}=\frac{\partial^2}{\partial x \partial y}u_n$. The second class consists of equations of the semi-discrete Toda-type lattice
\begin{equation}  \label{sdtodatype}
u_{n,x}^{j+1} = g( u_{n,x}^j, u_{n}^{j+1}, u_{n+1}^j, u_{n}^j,u_{n-1}^{j+1}),
\end{equation}
where only one of the arguments is continuous and the other two are discrete. The third class contains completely discrete models of Hirota-Miwa type
\begin{equation}  \label{hirotatype}
u_{n,m+1}^{j+1} = h( u_{n,m+1}^j, u_{n,m}^{j+1}, u_{n+1,m+1}^j, u_{n,m}^j,u_{n-1,m}^{j+1}).
\end{equation}
A complete description of integrable representatives of classes of nonlinear equations that are crucial for applications is a highly relevant and important task.
The well-known symmetry approach developed and successfully applied by A.B. Shabat, A.V. Zhiber, V.V. Sokolov, A.V. Mikhailov, S.I. Svinolupov, R.I. Yamilov, V.E. Adler and others has proven itself to be an effective tool for classifying integrable models in 1+1 dimensions. This approach yielded complete lists of integrable equations belonging to broad classes of differential and differential-difference equations (see, for instance, \cite{AdlerShabatYamilov}-\cite{LeviYamilov}).
 

There are very few classification results of integrable systems in 3D in general. The problem of a complete description of integrable models of type (\ref{todatype})-(\ref{hirotatype}) remains unsolved. The standard symmetry approach loses its efficiency when applied to equations with three independent variables due to the nonlocal variables involved in higher symmetries, which makes them difficult to calculate (see \cite{Yamilov}). Therefore, there is a need to search for new methods for classifying equations based on the characteristic features of three-dimensional integrable models.

Our research demonstrates that all known integrable discrete equations of the form (\ref{todatype})-(\ref{hirotatype}) admit finite-field reductions, which are obtained by imposing certain special (locking) boundary conditions along one of the discrete directions. The locking boundary condition is characterized by the following properties, emphasizing its uniqueness and significance:
\begin{enumerate}
	\item[1)] it is generated by imposing a constraint compatible with the lattice in question and the entire hierarchy of its higher symmetries (for a boundary condition compatible with symmetries, see \cite{Gurel-Hab97}, \cite{Gurses-Gurel-Hab94});
	\item[2)] when imposed at two distinct points $n_1$ and $n_2$ of a discrete line, they reduce the discrete equation to three equations that are in no way related to each other: two semi-infinite lattices defined on the rays $(-\infty, n_1)$ and $(n_2,\infty)$ and a finite-field hyperbolic system defined on the segment $(n_1, n_2)$;
	\item[3)] the resulting finite-field hyperbolic system has complete sets of characteristic integrals in each of the characteristic directions, i.e., it is integrable in the Darboux sense.
\end{enumerate}

The first example of the special (locking) boundary  condition for three-dimensional lattices can be found in the works of Laplace, where he proposed a method for solving a linear equation of the form:
\begin{equation}\label{linear-eq}
u_{xy}+a(x,y)u_x+b(x,y)u_x +c(x,y)u=0.
\end{equation}
Laplace considered differential substitutions, transforming the equation \eqref{linear-eq} into an equation of the same form. Under iterations of these substitutions, the Laplace invariants $v_n$ of the resulting linear equations satisfy lattice $E_2$ in the list below:
\begin{equation}\label{Laplace-inv}
(\ln v_n)_{xy}=v_{n+1}-2v_{n}+ v_{n-1}.
\end{equation}
In the Laplace method, the decisive role is played by the cut-off conditions of the form $v_{n_0}=0$, associated with the singular point of the equation \eqref{Laplace-inv}. For example, if we assume that $v_{0}=0$ and $v_{n_{N+1}}=0$, we obtain three independent hyperbolic systems defined for values of $n$ in the sets $(-\infty, -1]$, $[1,N]$, and $[N+2,+\infty)$. The finite-field system defined on the interval $[1,N]$ is Darboux integrable; more precisely, it can be solved explicitly (for more details see preprint \cite{Ganzha}).

It should be noted that Darboux integrable finite-field reductions of integrable three-dimensional lattices have applications in mathematical physics. For example, they can serve as an effective tool for constructing new solutions of nonlinear integrable equations in 3D (see, e.g., \cite{HabKh-24Umj}, \cite{GarHab-25PhysD}, \cite{HabKh-26RChD}). 

In the paper \cite{GarHab-25JPA} a two-stage algorithm for the identification and classification of integrable nonlinear lattices in 3D is proposed, also based on Darboux-integrable finite-field reductions followed by the application of higher symmetries.

In this paper, we use finite-field reductions of integrable lattices of class (\ref{sdtodatype}) to search for their discrete versions of the form (\ref{hirotatype}). The search for a lattice (\ref{hirotatype}) given an integrable lattice of the form (\ref{sdtodatype}) is performed by using complete sets of characteristic integrals of a special finite-field reduction of equation (\ref{sdtodatype}). Based on the examples discussed in this paper, it is easy to see that discretization using characteristic integrals is also discretization in the generally accepted sense. With an appropriate choice of the step parameter $\varepsilon$, the resulting discrete models transform into corresponding semi-discrete analogs in the continuum limit with $\varepsilon\rightarrow0$.

The discretization method we use, based on characteristic integrals, is described in detail below in Section 3 using the Liouville equation as an example (see also \cite{HabZhSak-11JMP}).

In paper \cite{Habibullin2013}, the following conjecture was formulated. 

{\bf Conjecture.} {\it Any integrable lattice of type (\ref{todatype}) admits finite-field reductions in the form of hyperbolic systems of equations of arbitrary order that admit complete sets of independent characteristic integrals, i.e., are Darboux-integrable systems. These reductions are obtained by imposing special boundary conditions on the lattice, such as truncation with respect to a discrete variable $n$; more precisely, they have the form:
\begin{equation}\label{fr1}
\begin{cases}
u_{1,xy} = f_1(u_{2},u_{1}, u_{1,x},u_{1,y}),\\
u_{n,xy} = f(u_{n+1},u_{n},u_{n-1}, u_{n,x},u_{n,y}),\quad 2\leq n \leq N-1,\\
u_{N,xy} = f_N(u_{N},u_{N-1}, u_{N,x},u_{N,y}).
\end{cases}
\end{equation}}
It should be noted that for the Toda lattice, Darboux-integrable reductions, called open chains in some sources, were studied back in the 19th century in the works of Laplace, Darboux, Goursat, Moutard and others.

It is well known that integrability in the Darboux sense has an equivalent algebraic interpretation based on the concept of the characteristic algebra of a hyperbolic system introduced by A.B. Shabat in 1980. A hyperbolic system is Darboux integrable if and only if its characteristic algebras in both directions have finite dimension. 

The above-mentioned conjecture was convincingly confirmed by all known examples of integrable lattices of type (\ref{todatype}) and tested as a classification criterion in the works \cite{PoptsovaHabibullin18}, \cite{Kuznetsova19}, \cite{HabibullinKuznetsova20}. The listed articles solve the problem of describing all integrable lattices of the form (\ref{todatype}) in the quasilinear case, i.e. with the additional requirement that the function $f(u_{n+1},u_{n},u_{n-1}, u_{n,x},u_{n,y})$ depends linearly on the variables $u_{n,x}$ and $u_{n,y}$.
We present a list of all currently known integrable Toda-type lattices, up to point transformations:
\begin{itemize}
\item[$E_1)$] $u_{n,xy} = e^{u_{n+1} - 2 u_n + u_{n-1} }$;
\item[$E_2)$] $u_{n,xy} = e^{u_{n+1}} - 2 e^{u_n} + e^{u_{n-1}}$;
\item[$E_3)$] $u_{n,xy} = e^{u_{n+1}-{u_n}} -  e^{u_n-u_{n-1}}$;
\item[$E_4)$] $u_{n,xy} = \left(u_{n+1} - 2 u_n + u_{n-1}  \right) u_{n,x}$;
\item[$E_5)$] $u_{n,xy} = \left(e^{u_{n+1}-{u_n}} -  e^{u_n-u_{n-1}}\right)u_{n,x}$;
\item[$E_6)$] $u_{n,xy}=\alpha_nu_{n,x}u_{n,y}, \quad \alpha_n = \frac{1}{u_n - u_{n-1}} - \frac{1}{u_{n+1}-u_n}$.
\end{itemize}
Lattices $E_1$–$E_3$ were obtained and studied in the classical works of Laplace, Darboux, Goursat,  Moutard and others. Lattices $E_4$ and $E_5$ apparently appeared in the paper \cite{ShabatYamilov97}. Equation $E_6$ was obtained simultaneously and independently in \cite{ShabatYamilov97} and \cite{Ferapontov97}. A discussion of the irreversible substitutions connecting the equations of the list can be found in the paper \cite{ShabatYamilov97}.

In the works \cite{HabibullinKhakimova21}, \cite{HabibullinKhakimovaTMP22}, \cite{HabibullinKhakimovaUMJ22}, \cite{KhakimovaLJM24} versions of the above conjecture suitable for lattices (\ref{sdtodatype}) and (\ref{hirotatype}) were studied.
It was observed that all integrable models belonging to the classes (\ref{sdtodatype}) and (\ref{hirotatype}) admit Darboux integrable reductions of the following types: 
\begin{equation}\label{gr1}
\begin{cases}
u_{1,x}^{j+1} = g_1( u_{1,x}^j, u_{1}^{j+1}, u_{2}^j, u_{1}^j),\\
u_{n,x}^{j+1} = g( u_{n,x}^j, u_{n}^{j+1}, u_{n+1}^j, u_{n}^j,u_{n-1}^{j+1}),\quad 2\leq n \leq N-1,\\
u_{N,x}^{j+1} = g_N( u_{N,x}^j, u_{N}^{j+1}, u_{N}^j,u_{N-1}^{j+1}),
\end{cases}
\end{equation}
and, respectively,
\begin{equation}\label{hr1}
\begin{cases}
u_{1,m+1}^{j+1} = h_1( u_{1,m+1}^j, u_{1,m}^{j+1}, u_{2,m+1}^j, u_{1,m}^j),\\
u_{n,m+1}^{j+1} = h( u_{n,m+1}^j, u_{n,m}^{j+1}, u_{n+1,m+1}^j, u_{n,m}^j,u_{n-1,m}^{j+1}),\quad 2\leq n \leq N-1,\\
u_{N,m+1}^{j+1} = h_N( u_{N,m+1}^j, u_{N,m}^{j+1}, u_{N,m}^j,u_{N-1,m}^{j+1}).
\end{cases}
\end{equation}

In this paper, we apply the discretization algorithm due to characteristic integrals presented in Section 3 below to derive integrable models of type \eqref{hirotatype}. To this end we use the list of equations of type \eqref{sdtodatype}, that consists of the following models 
\begin{enumerate}
	\item[$L_1$)] $u_{n,x}^{j+1}=u^j_{n,x}+e^{u^j_n-u^{j+1}_{n-1}}-e^{u_{n+1}^j-u_n^{j+1}};$
	\item[$L_2$)] $u_{n,x}^{j+1}=u_{n,x}^{j}-e^{u_{n-1}^{j+1}-u_n^{j}}+e^{u_{n}^{j+1}-u_{n+1}^{j}}-e^{u_{n-1}^{j+1}-u_n^{j+1}}+e^{u_{n}^{j}-u_{n+1}^{j}};$
	\item[$L_3$)] $u_{n,x}^{j+1}=u_{n,x}^{j}\frac{\left(u_n^{j+1}\right)^2}{u_{n+1}^{j}u_{n-1}^{j+1}};$
	\item[$L_4$)] $u_{n,x}^{j+1}=u_{n,x}^{j}\frac{u_{n+1}^{j}-u_n^{j+1}}{u_n^{j}-u_{n-1}^{j+1}};$
	\item[$L_5$)] $u_{n,x}^{j+1}=u_{n,x}^{j}\frac{u_n^{j+1}\left(u_n^{j+1}-u_{n+1}^{j}\right)}{u_{n+1}^{j}\left(u_{n-1}^{j+1}-u_n^{j}\right)};$
	\item[$L_6$)] $u^{j+1}_{n,x}=u^j_{n,x}\frac{\left(u^{j+1}_{n}-u^{j+1}_{n-1}\right)\left(u^{j+1}_{n}-u^{j}_{n+1}\right)}{\left(u^{j}_{n+1}-u^{j}_{n}\right)\left(u^{j+1}_{n-1}-u^{j}_{n}\right)};$
	\item[$L_7$)] $u^{j+1}_{n,x}=u^j_{n,x}+e^{u_{n-1}^{j+1}-u_n^{j+1}-u^j_n+u_{n+1}^j};$ 
	\item[$L_8$)] $u_{n,x}^{j+1}=u_{n,x}^{j}+e^{u_{n+1}^{j}}+e^{u_{n-1}^{j+1}}-e^{u_{n}^{j+1}}-e^{u_{n}^{j}};$ 
	\item[$L_9$)] $u_{n,x}^{j+1}=u_{n,x}^{j}+\left(u_n^{j+1}-u_n^{j}\right)\left(u_n^{j+1}+u_n^{j}-u_{n-1}^{j+1}-u_{n+1}^{j}\right).$
\end{enumerate}

Let us briefly discuss on the content of the article. In \S 2 we recall the notion of a complete set of characteristic integrals for hyperbolic systems of three kinds of models: partial differential equations, differential-difference lattices and fully discrete lattices. Here we formulate also a criterion of a complete set of integrals. In \S 3 we explain the method of discretization by preserving of the complete sets of characteristic integrals. The idea of the method is illustrated using the Liouville equation as an example. A rigorous algorithm  for deriving reductions  integrable in the sense of Darboux  is shown in \S 4. In \S5-\S13 we applied the method of discretization given in \S 3 to the semi-discrete Toda-type lattices $L_1$-$L_9$. In \S 14 the list of Hirota-Miwa type lattices obtained due to the discretization is discussed. The associated linear systems (Lax pairs) for these models are presented.

\pagebreak

\section{Complete sets of characteristic integrals}

\subsection{Hyperbolic type systems of PDE}

In this section we discuss the problem of constructing characteristic integrals for hyperbolic systems of the form
\begin{equation}\label{system}
u_{i,xy}=F_i(u,u_x, u_y), \quad i=1,2,\dots, N,
\end{equation}
where $u=(u_1,u_2,\dots, u_N)$, $u_x=(u_{1,x},u_{2,x},\dots, u_{N,x})$, $u_y=(u_{1,y},u_{2,y},\dots, u_{N,y})$.
It is supposed that functions $F_i$ are defined and analytic in some domain of complex space $\mathbb{C}^{3N}$.
For convenience we introduce the following notation $u_{[r]}:=\frac{\partial^r}{\partial x^r}u$.

 A smooth function $I=I(x,y,u, u_{[1]}, u_{[2]},...,u_{[r]})$ is called a $y$-integral of the order $r$ for \eqref{system} if it satisfies the condition $D_yI=0$, where $D_y$ denotes the total derivative operator with respect to $y$. In a similar way one defines characteristic integrals in the direction of $x$. Functions of the form $I(x)$ and $J(y)$ are called trivial integrals. System \eqref{system} is called integrable in sense of Darboux  if it admits a complete set of nontrivial functionally independent integrals in each of its characteristic directions $x$ and $y$. Completeness of the set of  characteristic integrals means that any $y$-integral (any $x$-integral) can be represented as a function of the integrals in the set and of their derivatives  with respect to $x$ (with respect to $y$).

Let us now briefly discuss the method of finding characteristic integrals. In this article, we restrict ourselves to the autonomous case, since in our subsequent research we will use only autonomous integrals -- that is, integrals that do not depend explicitly on $x$ and $y$.

Assume that $I=I( u, u_{[1]},u_{[2]},\ldots,u_{[k]})$ is a non-trivial $y$-integral for system (\ref{system}).
We rewrite equation $D_yI=0$ in expanded form
\begin{equation}\label{oper}
D_yI=\sum_{i=1}^N\left(u_{i,y}\frac{\partial}{\partial{u_i}}+F_{i}\frac{\partial}{\partial{u_{i,[1]}}}+D_x(f_i)\frac{\partial}{\partial{u_{i,[2]}}}+\cdots \right)I=0,
\end{equation}
where $u_{i,[k]}=D^k_xu_{i}$. Equation \eqref{oper} decomposes into the system of relations
\begin{equation}\label{x2}
\begin{cases}
X_iI=0,\quad i=1,2,\dots, N,\\
YI=0,
\end{cases}
\end{equation}
where
\begin{equation}\label{x3}
\begin{aligned}
&X_i=\frac{\partial}{\partial{u_{i,y}}},\quad i=1,2,\dots, N,\\
&Y= \sum^N_{i=1}\left(F_i \frac{\partial}{\partial{u_{i,[1]}}} +D_x(F_i) \frac{\partial}{\partial{u_{i,[2]}}}+\dots +D_x^{r-1}(F_i) \frac{\partial}{\partial{u_{i,[r]}}}\right),
\end{aligned}
\end{equation}
  because the variables $u_{i,y}$ appearing in the equation can be treated as independent parameters.

Therefore any autonomous $y$-integral is a solution of the system \eqref{x2}. To investigate the solvability conditions for the system \eqref{x2}, we employ the concept of the characteristic Lie algebra (see \cite{ShabatYamilov81,{Habibullin2013}, {PoptsovaHabibullin18},{HabibullinKhakimova21},{HabibullinKhakimovaTMP22},{Millionshchikov}}). Let us recall its definition. Assume that $A$ denotes the ring of locally analytic functions of the dynamical variables
$u_y,u, u_{[1]}, u_{[2]},\dots$. We consider the Lie algebra $L_y$ with the multiplication defined as $[H_1,H_2]=H_1H_2-H_2H_1$ generated by the set of the operators in \eqref{x3} over the ring $A$. We suppose  that additional  consistency conditions:
\begin{itemize}
\item[1)] $[H_1,aH_2]=H_1(a)H_2+a[H_1,H_2]$,
\item[2)] $(aH_1)b=aH_1(b)$
\end{itemize} 
are met for any $H_1,H_2\in L_y$ and $a,b\in A$.

Algebra $L_y$ is called characteristic Lie algebra in the direction $y$. Characteristic Lie algebra $L_x$ in the direction $x$ is defined in a similar way. Algebra $L_y$ is finite dimensional if it admits a finite basis consisting of the operators $Y_1, Y_2, \dots, Y_k\in L_y$ such that any element $Y\in L_y$ can be represented as follows
$$ Y=a_1Y_1+a_2 Y_2+ \dots+ a_kY_k$$
with the coefficients $a_1,a_2,\dots,a_k\in A$.

According to the definition of the characteristic algebra an arbitrary $y$-integral  belongs to the kernel of any operator in $L_y$.

Below we give a convenient tool for constructing a basis of the finite dimensional characteristic algebra and formulate a criterion of finite dimensionality.

{\bf Remark 1.} \textit{Following A.V. Zhiber \cite{KostriginaZhiber11} (see also \cite{ZMHS}), we introduce the notion of the complete set
of non-trivial integrals of minimal orders $0<r_1 < r_2 < \cdots < r_N.$ The minimality of the order $r_1$ means that the function
of the form
$$
w = w(x, y, u, u_{x},u_{xx},\dots,u_{[k]}),
$$
satisfying the condition $D_y(w) = 0$ and having the order $k$ less than $r_1$ is a trivial integral.
In this case, evidently, there is at least one non-trivial $y$-integral of order~$r_1$. We let $w_{1,1}, w_{1,2},\dots, w_{1,m_1}$ denote the complete set of functionally
independent $y$-integrals of order $r_1$. Next, we choose the number $r_2$ such that any $y$-integral of order $k<r_2$
is a function of $x$ and of the $y$-integrals $w_{1,1},\dots,w_{1,m_1}$ constructed at the preceding step, and also of their derivatives with respect to $x$. But at $k = r_2$, there are essentially new $y$-integrals, i.e., $y$-integrals that cannot be expressed in
terms of the $y$-integrals found previously. We let $w_{2,1},\ldots,w_{2,m_2}$ denote the set of functionally independent
$y$-integrals of order $r_2$ that cannot be expressed in terms of $x$, the $y$-integrals $w_{1,1},\ldots,w_{1,m_1}$, and their shifts.
Continuing this process, we obtain the set of $y$-integrals of minimal orders. Similarly, we can define the set
of $x$-integrals of minimal orders.}

The set of $y$-integrals $\left\{I_l\right\}^{l=N}_{l=1}$, having minimal orders $r_l$ for (\ref{system}), constitutes a complete set of independent $y$-integrals iff the condition is satisfied (see \cite{ZMHS})
\begin{equation}  \label{x6}
\mathrm{det}\,\left( \frac{\partial I_l}{\partial u^i_{n,[r_l]}} \right):=
\left|  
\begin{array}{cccc}
\frac{\partial I_1}{\partial u_{1,[r_1]}} & \frac{\partial I_1}{\partial u_{2,[r_1]}} & \dots & \frac{\partial I_1}{\partial u_{N,[r_1]}} \\
\dots & \dots & \dots & \dots\\
\frac{\partial I_N}{\partial u_{1,[r_N]}} & \frac{\partial I_N}{\partial u_{2,[r_N]}} & \dots & \frac{\partial I_N}{\partial u_{N,[r_N]}} 
\end{array}
 \right| \neq 0.
\end{equation}

For the set of $x$-integrals for (\ref{system}) we have a similar statement. In what follows we use inequalities of the form (\ref{x6}) to verify the Darboux integrability property of hyperbolic type systems.

{\bf Theorem 1.} {\it System of equations (\ref{system})
admits a complete set of independent autonomous  $y$-integrals if and only if the characteristic Lie algebra $L_y$ over the ring of locally analytic functions of the variables $ u_y$, $ u$, $ u_x$, $ u_{xx}$, $\ldots$, where $u=(u_1,u_2,\ldots, u_N)$, generated by the characteristic operators $X_i,\quad i=1,2,\dots N$ and $Y$ defined in \eqref{x3}
\begin{align*}
&X_i=\frac{\partial}{\partial u_{i,y}},\quad i=1,\ldots, N;\\
&Y=\sum_{i=1} u_{i,y}\frac{\partial}{\partial u_i}+ F_i \frac{\partial}{\partial u_{i,x}}+ D_x(F_i)\frac{\partial}{\partial u_{i,xx}}+\cdots,
\end{align*}
satisfies the following requirement. The dimension of $L_y$, as a left module over the ring of locally analytic functions, is finite.}

Let us consider Liouville equation
\begin{equation}\label{Liu}
u_{xy}=e^u
\end{equation}
as an illustrative example. Lie algebra for \eqref{Liu} is generated by the operators
\begin{equation}\label{XYLiu}
Z=\frac{\partial}{\partial u}, \quad W=e^u\frac{\partial}{\partial u_{x}}+ D_x(e^u)\frac{\partial}{\partial u_{xx}}+D^2_x(e^u)\frac{\partial}{\partial u_{xxx}}+\cdots.
\end{equation}
Evidently $[Z,W]=W$, therefore Lie algebra $L_y$ has dimension two, and operators $Z$ and $W$ form a basis for it (see \cite{{ZM06},{Millionshchikov}}).
Now we look for autonomous integrals of the minimal orders. According to the scheme above we start with the first order integral $I=I(u_x)$ satisfying the equation 
$$YI=e^u\frac{\partial}{\partial u_{x}}I(u_x)=0.$$ 
Then we observe that such an integral is trivial. Looking for autonomous integral of the second order $I=I(u_x, u_{xx})$
we get immediately $I=u_{xx}-\frac{1}{2}u_x^2$. Consequently, the complete set of independent integrals contains only one integral.

The concept of a characteristic algebra naturally extends to the case of semi-discrete and purely discrete systems of hyperbolic type. We do not present a detailed account of the properties of characteristic algebras in these cases but briefly state the definitions and main propositions (see below \S2.2, \S2.3). The reader may consult the works \cite{Habibullin2013}, \cite{HabibullinKhakimova21}, \cite{HabibullinKhakimovaUMJ22}, \cite{KhakimovaLJM24} for further details.

Below, we employ an algorithm based on the concept of a characteristic Lie algebra to construct complete sets of characteristic integrals for finite-field reductions of lattices $L_1$-$L_9$. A detailed account of the algorithm's application to systems of equations is provided in \S5, using a specific example of a finite-field reduction of lattice $L_1$.

Various methods for constructing characteristic integrals are discussed in \cite{Demskoi2010}-\cite{GurZh-04}. 


\subsection {Hyperbolic type systems of semi-discrete equations}

In a similar way we define integrals for semi-discrete hyperbolic system of the form
\begin{equation}\label{sdhyp}
u_{n,x}^{j+1} = F_n(u_{n,x}^j, u^j, u^{j+1}), \qquad n=1,2,\dots, N,
\end{equation}
where $u^j=(u_1^j, u_2^j, \dots, u_N^j)$. The system is related to the reduction (\ref{gr1}) of the semi-discrete Toda-type lattice.

A function
\begin{equation}\label{W}
W=W(x,j,u^{j-i},u^{j-i+1},\dots, u^{j+k}),
\end{equation}
where $i,k=0,1,2,\dots$, is called an $x$-integral of system (\ref{sdhyp}) of order $i+k$ if, for at least one pair of numbers $l,s=1,2,\dots,N$, the product $\frac{\partial W}{\partial u^{j-i}_l}\cdot \frac{\partial W}{\partial u^{j+k}_s}$ does not vanish identically and, in addition, the relation $D_xW=0$ holds. Here $D_x$ denotes the operator of total differentiation with respect to $x$. The derivative is calculated by means of system (\ref{sdhyp}). Since the operator $D_x$ commutes with the shift operator $D_j$, acting according to the rule $D_jf(j)=f(j+1)$, the operator $D_j$ maps the $x$-integral to the $x$-integral again. Therefore, below, when considering integrals, we will assume that  $i=0$ in the formula (\ref{W}).

Let us given a set of $x$-integrals

\begin{equation}\label{setW}
W_l=W_l(x,j,u^{j},u^{j+1},\dots, u^{j+k_l}),\quad l=1,2,\dots, N
\end{equation}
of the minimal orders $k_l$.

The set (\ref{setW}) constitutes a complete set  of $x$-integrals for the hyperbolic type system (\ref{sdhyp}) iff the condition 
\begin{equation}  \label{x7}
\mathrm{det}\,\left( \frac{\partial W_l}{\partial u^{j+k_l}_n} \right):=
\left|  
\begin{array}{cccc}
\frac{\partial W_1}{\partial u_1^{j+k_1}} & \frac{\partial W_1}{\partial u_2^{j+k_1}} & \dots & \frac{\partial W_1}{\partial u_N^{j+k_1}} \\
\dots & \dots & \dots & \dots\\
\frac{\partial W_N}{\partial u_1^{j+k_N}} & \frac{\partial W_N}{\partial u_2^{j+k_N}} & \dots & \frac{\partial W_N}{\partial u_N^{j+k_N}} 
\end{array}
 \right| \neq 0
\end{equation}
holds \cite{HabibullinKuznetsova21}.

A function of the form
\begin{equation*}\label{I}
I=I(x,j,u^{j},u^{j}_x, u^{j}_{xx},\dots, u^{j}_{[m]})
\end{equation*}
is called a $j$-integral of the order $m$ for system (\ref{sdhyp}) if the relation holds $D_j I=I$, where the shift of $j$ is evaluated due to the system (\ref{sdhyp}). Let us recall, that $u_{[m]}=\frac{\partial^m u}{\partial x^m}$.

Now we concentrate on a set of $j$-integrals
\begin{equation}\label{setI}
I_s=I(x,j,u^{j},u^{j}_x, u^{j}_{xx},\dots, u^{j}_{[m_s]}),\quad s=1,2,\dots, N
\end{equation}
of the minimal orders $m_s$. The set (\ref{setI}) defines a complete set of $j$-integrals for the system (\ref{sdhyp}) iff the condition 
\begin{equation*}  \label{x8}
\mathrm{det}\,\left( \frac{\partial I_s}{\partial u^{j}_{[m_s]}} \right):=
\left|  
\begin{array}{cccc}
\frac{\partial I_1}{\partial u^{j}_{1,[m_1]}} & \frac{\partial I_1}{\partial u^{j}_{2,[m_1]}} & \dots & \frac{\partial I_1}{\partial u^{j}_{N,[m_1]}} \\
\dots & \dots & \dots & \dots\\
\frac{\partial I_N}{\partial u^{j}_{1,[m_N]}} & \frac{\partial I_N}{\partial u^{j}_{2,[m_N]}} & \dots & \frac{\partial I_N}{\partial u^{j}_{N,[m_N]}} 
\end{array}
 \right| \neq 0
\end{equation*}
holds (see \cite{HabibullinKuznetsova21}).

\subsection {Hyperbolic type systems of fully discrete equations}

Let us consider the third type of hyperbolic systems
\begin{equation}\label{fdhyp}
u_{n,m+1}^{j+1} = F_n(u_{m}^j, u^j_{m+1}, u_m^{j+1}), \quad n=1,2,\dots, N,
\end{equation}
where $u^j_m=(u_{1,m}^j,u_{2,m}^j,\dots,u_{N,m}^j)$. 
It is associated with the reduction
(\ref{hr1}) of a fully discrete Hirota-Miwa type model (\ref{hirotatype}).
The system is assumed to be of hyperbolic type. This means that the system can be uniquely represented in each of the following three forms:
\begin{equation*}\label{hr-1}
\begin{cases}
u_{n,m-1}^{j+1} = F_n^{1,-1}( u_{m}^j, u_{m}^{j+1}, u_{m-1}^j),\\
u_{n,m+1}^{j-1} = F_n^{-1,1}( u_{m}^j, u_{m}^{j-1}, u_{m+1}^j),\\
u_{n,m-1}^{j-1} = F_n^{-1,-1}( u_{m}^j, u_{m}^{j-1}, u_{m-1}^j).
\end{cases}
\end{equation*}
A function of the form
\begin{equation}\label{J}
J=J(u_m^{j-k}, u_m^{j-k+1},\dots,u_m^{j+s}),\quad \mbox{where}\quad k,s\geq0
\end{equation}
is called an $m$-integral of the order $k+s$ for system (\ref{fdhyp}), if there exists a pair of numbers $n_1,n_2=1,2,\dots ,N$, such that the product 
$\frac{\partial J}{\partial u^{n+s}_{n_1,m}}\cdot \frac{\partial J}{\partial u^{n-k}_{n_2,m}}$ does not vanish identically and the following relation $D_mJ=J$ holds. Here $D_m$ stands for the operator shifting the argument $m$: $D_mf(m)=f(m+1)$. Since the shift operators $D_m$ and $D_j$ commute, the operator $D_j$ maps any $m$-integral to an $m$-integral. Therefore, without loss of generality, we can set $k=0$ in definition (\ref{J}).

Let us take a set of $m$-integrals 
\begin{equation}\label{setJ}
J_i=J_i(u_m^{j}, u_m^{j+1},\dots,u_m^{j+n_i}), \qquad i=1,2, \dots, N,
\end{equation}
of the minimal orders $n_1, n_2, \dots ,n_N$, constructed due to the rule discussed in \S 2.1.

The set (\ref{setJ}) constitutes a complete set of independent $m$-integrals if and only if the following relation 
\begin{equation*}  \label{x9}
\mathrm{det}\,\left( \frac{\partial J_i}{\partial u^{j}_{m}} \right):=
\left|  
\begin{array}{cccc}
\frac{\partial J_1}{\partial u^{j+n_1}_{1,m}} & \frac{\partial J_1}{\partial u^{j+n_1}_{2,m}} & \dots & \frac{\partial J_1}{\partial u^{j+n_1}_{N,m}} \\
\dots & \dots & \dots & \dots\\
\frac{\partial J_N}{\partial u^{j+n_N}_{1,m}} & \frac{\partial J_N}{\partial u^{j+n_N}_{2,m}} & \dots & \frac{\partial J_N}{\partial u^{j+n_N}_{N,m}} 
\end{array}
 \right| \neq 0
\end{equation*}
holds (see \cite{HabibullinKhakimovaTMP22}).

\section {Algorithm for discretization of lattices in 3D using integrals}
\subsection {Discussion of the algorithm in 1+1 dimensions}

Let us consider the rule for finding a differential-difference equation for a given integral of a partial differential equation using a simple example. As a touchstone, we will take the Liouville equation
\begin{equation}\label{L}
u_{xy}=e^u.
\end{equation}
As is known, the Liouville equation has characteristic integrals of the form
\begin{equation}\label{Li}
I=u_{xx}-\frac{1}{2}u_x^2, \qquad J=u_{yy}-\frac{1}{2}u_y^2.
\end{equation}
It is easy to verify that the equalities $D_yI=0$, $D_xJ=0$ hold, where the derivatives $D_y$ and $D_x$ are calculated by virtue of the equation (\ref{L}). We assume that the dynamical variables $u$, $u_x$, $u_{xx}$, $\ldots$ depend on an additional discrete $n$. By discretization of the equation (\ref{L}) using the y-integral (\ref{Li}) we mean an equation of the form
\begin{equation}\label{DL}
u_{n+1,x}=H(u_{n,x}, u_{n+1}, u_{n})
\end{equation}
such that the function 
\begin{equation}\label{Libar}
\bar I^{(n)}=u_{n,xx}-\frac{1}{2}u_{n,x}^2
\end{equation}
is an integral in the direction $n$ for this equation. In other words, the relation
\begin{equation}\label{DisI}
D_n\bar I^{(n)}=\bar I^{(n)}
\end{equation}
holds, where the shift is due to the equation (\ref{DL}). It is easy to verify (see \cite{HabZhSak-11JMP}) that function $H$ is uniquely determined from  condition (\ref{DisI}). The desired equation has the form (see, for instance, \cite{Zabrodin98}, \cite{AdleStartsev99})
\begin{equation}\label{DLE}
u_{n+1,x}=u_{n,x}+Ce^{\frac{1}{2} (u_{n+1}+ u_{n})},
\end{equation}
where $C$ is an arbitrary constant. 

{\bf Remark 2.} \textit{Equations (\ref{L}) and (\ref{DLE}) have the same characteristic integral (\ref{Libar}), since the relations $D_y\bar I^{(n)}=0$ and $D_n\bar I^{(n)}=\bar I^{(n)}$ hold. Therefore, the mapping that transforms the Liouville equation into (\ref{DLE}) can be called a discretization with preservation of the characteristic integral.}

Note that  equation (\ref{DLE}) admits an integral in another direction as well
\begin{equation}\label{Yi}
J=e^{\frac{1}{2} (u_{n+1}- u_{n})}+e^{\frac{1}{2} (u_{n+1}- u_{n+2})}.
\end{equation}
To verify this, we need to check that relation $D_xJ=0$ holds, where the derivative is taken by virtue of the equation (\ref{DLE}). The example shows that discretization by means of the integral preserves integrability in the sense of Darboux.

The next step in our discretization scheme is to find a fully discrete equation of the form
\begin{equation}\label{DDLE}
u_{n+1,m+1}=f(u_{n,m}, u_{n+1,m}, u_{n,m+1}),
\end{equation}
 by requiring that $x$-integral (\ref{Yi}) of the semi-discrete version of the Liouville equation (\ref{DLE}) generates an $m$-integral for the lattice (\ref{DDLE}). More precisely we look for equation  (\ref{DDLE}) for which the following function 

\begin{equation}\label{barJ}
\bar J=e^{\frac{1}{2} (u_{n+1,m}- u_{n,m})}+e^{\frac{1}{2} (u_{n+1,m}- u_{n+2,m})}
\end{equation}
would satisfy the relation
\begin{equation}\label{m-integral}
D_m\bar J=\bar J,
\end{equation}
where the function $D_m\bar J$ is evaluated by means of the equation (\ref{DDLE}).

To simplify the calculations, we replace the variables by setting $w_{n,m}=e^{-\frac{1}{2}u_{n,m}}$. In the new variables, the equation (\ref{DDLE}) and the integral (\ref{barJ}) take the forms, respectively,
\begin{equation}\label{DDLE2}
w_{n+1,m+1}=g(w_{n,m}, w_{n+1,m}, w_{n,m+1})
\end{equation}
and
\begin{equation*}\label{barJ2}
\bar J=\frac{w_{n,m}+ w_{n+2,m}}{ w_{n+1,m}}.
\end{equation*}
To look for  (\ref{DDLE2}) we use equation (\ref{m-integral}) that implies
\begin{equation}\label{main}
\frac{w_{n,m+1}+g(w_{n+1,m}, w_{n+2,m}, w_{n+1,m+1})}{g(w_{n,m}, w_{n+1,m}, w_{n,m+1})}=\frac{w_{n,m}+ w_{n+2,m}}{ w_{n+1,m}}.
\end{equation}
First, we differentiate the equation (\ref{main}) with respect to the variable $w_{n+2,m}$, and then apply the shift operator $D_n^{-1}$ to the result obtained. Finally, we arrive at the differential equation
$$
\frac{\partial g(w_{n,m}, w_{n+1,m}, w_{n,m+1})}{\partial w_{n+1,m}}=\frac{w_{n,m+1}}{w_{n,m}},
$$
which is easily integrated and leads to the following representation for $g$:
\begin{equation}\label{g6}
g=\frac{w_{n,m+1}w_{n+1,m}}{w_{n,m}}+c(w_{n,m},w_{n,m+1}),
\end{equation}
where the term $c(w_{n,m},w_{n,m+1})$ appears as a constant of integration. In order to find $c$ we substitute (\ref{g6}) into the main equation (\ref{main}). After simplification one gets the following relation
\begin{equation*}\label{eq7}
w_{n+1,m}c(w_{n+1,m},w_{n+1,m+1})=w_{n,m}c(w_{n,m},w_{n,m+1})
\end{equation*} 
that demonstrates that function $W:=w_{n,m}c(w_{n,m},w_{n,m+1})$ is an $n$-integral of the first order for the equation (\ref{DDLE2}). Since the equation is supposed to be a discretisation of the Liouville equation we assume that it does not admit any non-trivial first order integral. Therefore $W$ is a trivial autonomous integral of the form $W=C_1$, where $C_1$ is an arbitrary constant, i.e. we have $w_{n,m}c(w_{n,m},w_{n,m+1})=C_1$. Thus $c(w_{n,m},w_{n,m+1})=C_1/w_{n,m}$ and the desired equation is of the form
$$
w_{n+1,m+1}=\frac{ w_{n,m+1} w_{n+1,m}+C_1}{w_{n,m}}.
$$
Returning to the original variables, we arrive at the well-known fully discrete version of the Liouville equation:
\begin{equation*}\label{DDLiouville}
e^{-\frac{1}{2}(u_{n+1,m+1}+u_{n,m})}=e^{-\frac{1}{2}(u_{n+1,m}+u_{n,m+1})}+C_1
\end{equation*}
(see, for instance, \cite{Zabrodin98}, \cite{AdleStartsev99}).

\subsection{Scheme of discretization  in 3D.}

In this section, we elucidate the structure of discretization maps -- that is, maps that transform an integrable chain with three independent variables into its discrete analogue while preserving the set of integrals associated with a specific finite reduction. The main difference from the 1+1 case is that here we use not the chain itself, but a special finite reduction of it.

\subsubsection{Transition from the continuous model to the semi-discrete one.}

In our recent article \cite{HabKh-25MatSb} we have suggested an algorithm for discretising integrable Toda-type lattices (\ref{todatype}) based on use of the Darboux-integrable finite-field reductions (\ref{fr1}) of the lattices. The method for deriving reductions of the form (\ref{fr1}) can be found  in the work \cite{HabibullinKuznetsova20}. Let us briefly explain the essence of the method. At first we construct a complete set of $y$-integrals (or, alternatively a complete set of $x$-integrals) for (\ref{fr1})
\begin{equation}\label{setintegrals1}
I_l=I_l(x,y,u,u_{[1]}, u_{[2]}\dots, u_{[r_l]}), \quad l=1,2,\dots, N,
\end{equation}
where $u=(u_1,u_2,\dots u_N)$, $u_{[i]}=\frac{\partial}{\partial x^i}u$, $D_yI_l=0$. Afterwards we assume that the multicomponent variable $u=u(x,y)$ depends in addition on an upper index $j$, such that $u=u^j(x,y)$. Therefore the set (\ref{setintegrals1}) generates a new set of functions 
\begin{equation}\label{newsetintegrals1}
\bar I_l=\bar I_l(x,y,u^j,u_{[1]}^j, u_{[2]}^j\dots, u_{[r_l]}^j), \quad l=1,2,\dots, N.
\end{equation}
 The next step of our algorithm is to find a system of the form (\ref{gr1}), assuming that (\ref{newsetintegrals1}) provides a set of $j$-integrals for this system. In other words we require that the relations
\begin{equation}\label{jshift}
D_j\bar I_l=\bar I_l, \quad l=1,2,\dots, N
\end{equation}
be satisfied if the shift of the argument $j$ is calculated using (\ref{gr1}).
In fact, the relations (\ref{jshift}) provide a system of equations for determining unknown functions $g$, $g_1$, $g_N$ appearing in (\ref{gr1}). The most surprising and mysterious aspect is that, for all known equations of the form (\ref{todatype}), all the objects listed above are uniquely determined. Clearly, given the function $g$ one can easily reconstruct the desired lattice of the form (\ref{sdtodatype}), which is a discretization of the Toda-type lattice. The examples convincingly confirm that the lattice obtained as a result of these manipulations is indeed a discretization in the generally accepted sense. In other words, by calculating the continuum limit for the resulting discretization in the standard sense, we recover the original chain.

 {\bf Remark 3.} \textit{The functions (\ref{newsetintegrals1}) form a complete set of characteristic integrals for both (\ref{fr1}) and (\ref{gr1}). Namely, $y$-integrals for (\ref{fr1}) and $j$-integrals for (\ref{gr1}). Therefore, the mapping proposed above, which takes (\ref{fr1}) to (\ref{gr1}), can be called a discretization of the hyperbolic system with preservation of the complete set of characteristic integrals.}

\subsubsection{Transition from the semi-discrete model to the discrete one.}

The method of discretization by preserving of the complete set of characteristic integrals can be applied to the semi-discrete Toda-type lattices as well.
Below, we will focus on the relationship between the second and third classes of integrable discrete models (see, (\ref{sdtodatype}) and (\ref{hirotatype})). In particular, we will discuss the problem of integrable discretization of nonlinear lattices of the form (\ref{sdtodatype}). In fact, we establish some correspondence between the classes of equations of the form (\ref{sdtodatype}) and (\ref{hirotatype}). For this purpose, we pass from the lattice of type (\ref{sdtodatype}) under consideration to its Darboux-integrable finite-field reduction  (\ref{gr1}). To find the discretization, we use the complete set (\ref{setW}) of $x$-integrals of the system (\ref{gr1}). We assume that the dynamical variables $u^j_n(x)$ depend on additional discrete variable $m$ such that the set of functions (\ref{setW}) takes the form
\begin{equation}\label{setWW}
\bar W_l=\bar W_l(x,j,u^{j}_m,u^{j+1}_m,\dots, u^{j+k_l}_m),\quad l=1,2,\dots, N.
\end{equation}
By analogy with the previous case, we are looking for a hyperbolic system of the form (\ref{hr1}), for which the set of functions (\ref{setWW}) is a set of characteristic $m$-integrals. Using the middle equation of the constructed system (\ref{hr1}), we reconstruct the desired Hirota-Miwa type chain.

In Section 5, we describe in detail the entire process of discretizing semi-discrete Toda-type lattices, using chain $L_1$ as an example. In the remaining sections (\S6-\S13), we omit the details of calculations, as they are analogous to those for example $L_1$, and provide only the results obtained during discretization. 
By performing similar calculations as in \S5, we established a correspondence between equations of classes (\ref{sdtodatype}) and (\ref{hirotatype}). Note that this correspondence is not one-to-one, certain equations of class (\ref{hirotatype}) correspond to several equations of class (\ref{sdtodatype}).

\section{Discussion on the algorithm for searching the special (locking) boundary conditions for lattices of the form (\ref{sdtodatype})}

Let us explain the algorithm for constructing Darboux-integrable reductions of integrable lattices using the example of the equation $L_1$. It is based on the invariance of the lattice under a point transformation of the form $u^j_n \rightarrow u^j_n +(n+j)C$, where $C$ is an arbitrary constant. First, we perform the following change of variables
\begin{align*}
\begin{aligned}
&u_n^j=\bar u_n^j\pm(n+j)\ln \varepsilon \quad\mbox{for}\quad \pm n>0,\\
&u_0^j=\bar u_0^j+(-1)^{j+j_0}(1-j)\ln \varepsilon,
\end{aligned}
\end{align*}
here $\varepsilon>0$ ang $j_0$ is an integer.

We note that, due to the above-mentioned shift invariance of  lattice $L_1$, it does not change when $n\notin [-1,1]$. However, for $n\in [-1,1]$ it is  changed.
For these values of $n$ we have
\begin{align*}
\begin{aligned}
& \bar u_{1,x}^{j+1}=\bar u^j_{1,x}+\varepsilon e^{\bar u^j_1-\bar u^{j+1}_{0}}-e^{\bar u_{2}^j-\bar u_1^{j+1}}, \\
& \bar u_{0,x}^{j+1}=\bar u^j_{0,x}+\varepsilon e^{\bar u^j_0-\bar u^{j+1}_{-1}}-\varepsilon e^{\bar u_{1}^j-\bar u_0^{j+1}}, \\
& \bar u_{-1,x}^{j+1}=\bar u^j_{-1,x}+e^{\bar u^j_{-1}-u^{j+1}_{-2}}-\varepsilon e^{\bar u_{0}^j-\bar u_{-1}^{j+1}},
\end{aligned}
\end{align*}
where parameter $j_0$ is chosen in such a way that $j_0=0$ for even values of  $j$ and $j_0=1$ for odd $j$.

It is clear that for $\varepsilon=0$ the obtained system is simplified. For example, we obtain a relation of the form $\bar u_{0,x}^{j+1}=\bar u^j_{0,x}$. It shows that function $\bar u^j_{0}$ is a free functional parameter. For simplicity, we set $\bar u^j_{0}=0$. Actually this constraint is nothing else but the special cut-off boundary condition. Indeed,  it decomposes the entire system  into two independent semi-infinite subsystems defined on the semi-axes $n\in(-\infty, -1]$ and $n\in[1,+\infty)$ (below we omit the bar over the letter):
\begin{align*}
\begin{aligned}
&  u_{-1,x}^{j+1}= u^j_{-1,x}+e^{ u^j_{-1}-u^{j+1}_{-2}},\\
& u_{n,x}^{j+1}=u^j_{n,x}+e^{u^j_n-u^{j+1}_{n-1}}-e^{u_{n+1}^j-u_n^{j+1}}, \quad\mbox{for}\quad n\leq -2
\end{aligned}
\end{align*}
and
\begin{align}\label{right}
\begin{aligned}
&  u_{1,x}^{j+1}= u^j_{1,x}-e^{ u_{2}^j- u_1^{j+1}},\, \\
&  u_{n,x}^{j+1}=u^j_{n,x}+e^{u^j_n-u^{j+1}_{n-1}}-e^{u_{n+1}^j-u_n^{j+1}},\, \quad\mbox{for}\quad n\geq 2. 
\end{aligned}
\end{align}
In order to get the finite-field reduction of $L_1$ we have to derive the second cut-off condition for  (\ref{right}). To this end we apply to (\ref{right}) the following change of the variables
\begin{align*}
\begin{aligned}
&u_n^j=\tilde u_n^j+(n+j)\ln \varepsilon \quad\mbox{for}\quad  n>4,\\
&u_n^j=\tilde u_n^j-(n+j)\ln \varepsilon \quad\mbox{for}\quad 1\leq n<4,\\
&u_4^j=\tilde u_4^j+(-1)^{j+j_0}(-3+j)\ln \varepsilon.
\end{aligned}
\end{align*}
Assume that the integer parameter $j_0$ is chosen in such a way that $j_0=0$ for $j=2k$ and $j_0=1$ for $j=2k+1$. Then it is easy to verify that with such a change of variables only three lattice equations (\ref{right}) change, corresponding to $n=3$, $n=4$, $n=5$. These equations take the form (for simplicity we omit the tilde above the letter):
\begin{align*}
\begin{aligned}
&  u_{5,x}^{j+1}= u^j_{5,x}+\varepsilon e^{ u^j_5- u^{j+1}_{4}}-e^{ u_{6}^j- u_5^{j+1}}, \\
&  u_{4,x}^{j+1}= u^j_{4,x}+\varepsilon e^{ u^j_4- u^{j+1}_{3}}-\varepsilon e^{ u_{5}^j- u_4^{j+1}}, \\
&  u_{3,x}^{j+1}= u^j_{3,x}+e^{ u^j_{3}- u^{j+1}_{2}}-\varepsilon e^{ u_{4}^j- u_{3}^{j+1}}.
\end{aligned}
\end{align*}
By setting $\varepsilon=0$ and then assuming $ u^j_{4}=0$ we arrive to two independent objects: a semi-infinite lattice 
\begin{align*}
\begin{aligned}
&  u_{5,x}^{j+1}= u^j_{5,x}-e^{ u_{6}^j- u_5^{j+1}},\, \\
&  u_{n,x}^{j+1}=u^j_{n,x}+e^{u^j_n-u^{j+1}_{n-1}}-e^{u_{n+1}^j-u_n^{j+1}},\, \quad\mbox{for}\quad n\geq 6 
\end{aligned}
\end{align*}
and the desired finite-field system
\begin{align*}
\begin{aligned}
& u_{1,x}^{j+1}=u^{j}_{1,x}-e^{u_2^j-u_1^{j+1}}, \\
& u_{2,x}^{j+1}=u^{j}_{2,x}+e^{u_2^j-u_1^{j+1}}-e^{u_3^j-u_2^{j+1}}, \\
& u_{3,x}^{j+1}=u^{j}_{3,x}+e^{u_3^j-u_2^{j+1}}.
\end{aligned}
\end{align*}
Let us remark that the algorithm is effective for the lattices when the right-hand side of the lattice has an essential singularity at the point $u=\infty$ (see $L_1$, $L_2$, $L_7$, $L_8$). For the integrable lattices depending rationally or polynomially on the variables $u^j_n$ appropriate boundary conditions are related to stationary solutions of the lattices.

\section{Discretization of lattice $L_1$}

In this section, for the sake of completeness, we present a detailed description of the procedure for deriving a fully discrete Hirota-Miwa  type analogue for the semi-discrete $L_1$ lattice
\begin{equation}\label{L1}
u_{n,x}^{j+1}=u^j_{n,x}+e^{u^j_n-u^{j+1}_{n-1}}-e^{u_{n+1}^j-u_n^{j+1}}
\end{equation}
which was previously developed in \cite{Khakimova-26}.  

The first step in our algorithm consists in finding a suitable finite reduction of equation \eqref{L1}
\begin{equation}\label{sysL1}
\begin{cases}
\displaystyle u_{1,x}^{j+1}=u^{j}_{1,x}-e^{u_2^j-u_1^{j+1}}, \\
\displaystyle u_{2,x}^{j+1}=u^{j}_{2,x}+e^{u_2^j-u_1^{j+1}}-e^{u_3^j-u_2^{j+1}}, \\
\displaystyle u_{3,x}^{j+1}=u^{j}_{3,x}+e^{u_3^j-u_2^{j+1}},
\end{cases} 
\end{equation}
the derivation of which was discussed in the previous section.

To apply the discretization method, we require the $x$-integrals of system \eqref{sysL1}. The $x$-integrals of system \eqref{sysL1} were previously found in \cite{Smirnov15} using a Lax representation; here, however, we employ the method of characteristic Lie algebras for this purpose. Therefore, at first we have to describe the characteristic Lie algebra.

We rewrite system \eqref{sysL1} in inverse form
\begin{equation}\label{sysL1-1}
\begin{cases}
\displaystyle u_{1,x}^{j-1}=u^{j}_{1,x}+e^{u_2^{j-1}-u_1^{j}}, \\
\displaystyle u_{2,x}^{j-1}=u^{j}_{2,x}-e^{u_2^{j-1}-u_1^{j}}+e^{u_3^{j-1}-u_2^{j}}, \\
\displaystyle u_{3,x}^{j-1}=u^{j}_{3,x}-e^{u_3^{j-1}-u_2^{j}}.
\end{cases} 
\end{equation}
Suppose that function $W(u_1^j,u_2^j,u_3^j,u_1^{j+1},u_2^{j+1},u_3^{j+1},u_1^{j-1},u_2^{j-1},u_3^{j-1},u_1^{j+2},\ldots)$ is an $x$-integral of system \eqref{sysL1}. Then, according to the definition, relation 
\begin{equation*}
D_xW(u_1^j,u_2^j,u_3^j,u_1^{j+1},u_2^{j+1},u_3^{j+1},u_1^{j-1},u_2^{j-1},u_3^{j-1},u_1^{j+2},\ldots)=0
\end{equation*}
holds. By applying the chain rule we rewrite the latter in the following form:
\begin{equation*}
K_0W=0,
\end{equation*}
where
\begin{align*}
K_0=&u_{1,x}^j\frac{\partial}{\partial u_1^j}+u_{2,x}^j\frac{\partial}{\partial u_2^j}+u_{3,x}^j\frac{\partial}{\partial u_3^j}+u_{1,x}^{j+1}\frac{\partial}{\partial u_1^{j+1}}+u_{2,x}^{j+1}\frac{\partial}{\partial u_2^{j+1}}+u_{3,x}^{j+1}\frac{\partial}{\partial u_3^{j+1}}\\
&+u_{1,x}^{j-1}\frac{\partial}{\partial u_1^{j-1}}+u_{2,x}^{j-1}\frac{\partial}{\partial u_2^{j-1}}+u_{3,x}^{j-1}\frac{\partial}{\partial u_3^{j-1}}+u_{1,x}^{j+2}\frac{\partial}{\partial u_1^{j+2}}+\ldots.
\end{align*}
Afterwards we simplify  operator $K_0$  by eliminating  variables $u_{1,x}^{j\pm1}$, $u_{2,x}^{j\pm1}$, $u_{3,x}^{j\pm1}$, $u_{1,x}^{j\pm2}, \ldots$ due to systems \eqref{sysL1} and \eqref{sysL1-1}. As a result we get 
\begin{align*}
K_0=&u_{1,x}^j\frac{\partial}{\partial u_1^j}+u_{2,x}^j\frac{\partial}{\partial u_2^j}+u_{3,x}^j\frac{\partial}{\partial u_3^j}+\left(u^{j}_{1,x}-e^{u_2^j-u_1^{j+1}}\right)\frac{\partial}{\partial u_1^{j+1}}\\
&+\left(u^{j}_{2,x}+e^{u_2^j-u_1^{j+1}}-e^{u_3^j-u_2^{j+1}}\right)\frac{\partial}{\partial u_2^{j+1}}+\left(u^{j}_{3,x}+e^{u_3^j-u_2^{j+1}}\right)\frac{\partial}{\partial u_3^{j+1}}\\
&+\left(u^{j}_{1,x}+e^{u_2^{j-1}-u_1^{j}}\right)\frac{\partial}{\partial u_1^{j-1}}+\left(u^{j}_{2,x}-e^{u_2^{j-1}-u_1^{j}}+e^{u_3^{j-1}-u_2^{j}}\right)\frac{\partial}{\partial u_2^{j-1}}\\
&+\left(u^{j}_{3,x}-e^{u_3^{j-1}-u_2^{j}}\right)\frac{\partial}{\partial u_3^{j-1}}+\left(u^{j}_{1,x}-e^{u_2^j-u_1^{j+1}}-e^{u_2^{j+1}-u_1^{j+2}}\right)\frac{\partial}{\partial u_1^{j+2}}+\ldots.
\end{align*}
Note that the coefficients of operator $K_0$ depend on variables $u^{j}_{1,x}$, $u^{j}_{2,x}$, $u^{j}_{3,x}$, whereas function $W$ does not depend on them. In other words, operators  
\begin{align*}
&X_1=\frac{\partial}{\partial u_1^j}+\frac{\partial}{\partial u_1^{j+1}}+\frac{\partial}{\partial u_1^{j-1}}+\frac{\partial}{\partial u_1^{j+2}}+\ldots,\\
&X_2=\frac{\partial}{\partial u_2^j}+\frac{\partial}{\partial u_2^{j+1}}+\frac{\partial}{\partial u_2^{j-1}}+\frac{\partial}{\partial u_2^{j+2}}+\ldots,\\
&X_3=\frac{\partial}{\partial u_3^j}+\frac{\partial}{\partial u_3^{j+1}}+\frac{\partial}{\partial u_3^{j-1}}+\frac{\partial}{\partial u_3^{j+2}}+\ldots
\end{align*}
also annihilate function $W$, that is, the following equalities hold:
\begin{align*}
X_1W=0, \qquad X_2W=0, \qquad X_3W=0. 
\end{align*}
Consequently, the operator $K_0$ can be represented as a linear combination
\begin{align*}
K_0=u^{j}_{1,x}X_1+u^{j}_{2,x}X_2+u^{j}_{3,x}X_3+K,
\end{align*}
where 
\begin{align*}
K=&-e^{u_2^j-u_1^{j+1}}\frac{\partial}{\partial u_1^{j+1}}+\left(e^{u_2^j-u_1^{j+1}}-e^{u_3^j-u_2^{j+1}}\right)\frac{\partial}{\partial u_2^{j+1}}+e^{u_3^j-u_2^{j+1}}\frac{\partial}{\partial u_3^{j+1}}\\
&+e^{u_2^{j-1}-u_1^{j}}\frac{\partial}{\partial u_1^{j-1}}+\left(-e^{u_2^{j-1}-u_1^{j}}+e^{u_3^{j-1}-u_2^{j}}\right)\frac{\partial}{\partial u_2^{j-1}}\\
&-e^{u_3^{j-1}-u_2^{j}}\frac{\partial}{\partial u_3^{j-1}}+\left(-e^{u_2^j-u_1^{j+1}}-e^{u_2^{j+1}-u_1^{j+2}}\right)\frac{\partial}{\partial u_1^{j+2}}+\ldots.
\end{align*}
Thus, we obtain the following system of equations for the unknown function~$W$:
\begin{align*}
X_1W=0, \qquad X_2W=0, \qquad X_3W=0, \qquad KW=0. 
\end{align*}
Let us check whether the characteristic algebra consisting of operators $X_1$, $X_2$, $X_3$, $K$ is finite. To do this, we compute all possible commutators of these operators. It is clear that the commutators of the operators $X_1$, $X_2$, $X_3$ are zero, i.e. we have $[X_1,X_2]=0$, $[X_1,X_3]=0$, $[X_2,X_3]=0$. Now we calculate $[X_1,K]$, $[X_2,K]$, $[X_3,K]$:
\begin{align*}
K_1:=&[X_1,K]=e^{u_2^j-u_1^{j+1}}\frac{\partial}{\partial u_1^{j+1}}-e^{u_2^j-u_1^{j+1}}\frac{\partial}{\partial u_2^{j+1}}-e^{u_2^{j-1}-u_1^{j}}\frac{\partial}{\partial u_1^{j-1}}\\
&+e^{u_2^{j-1}-u_1^{j}}\frac{\partial}{\partial u_2^{j-1}}+\left(e^{u_2^j-u_1^{j+1}}+e^{u_2^{j+1}-u_1^{j+2}}\right)\frac{\partial}{\partial u_1^{j+2}}+\ldots,\\
K_2:=&[X_2,K]=-e^{u_2^j-u_1^{j+1}}\frac{\partial}{\partial u_1^{j+1}}+\left(e^{u_2^j-u_1^{j+1}}+e^{u_3^j-u_2^{j+1}}\right)\frac{\partial}{\partial u_2^{j+1}}\\
&-e^{u_3^j-u_2^{j+1}}\frac{\partial}{\partial u_3^{j+1}}+e^{u_2^{j-1}-u_1^{j}}\frac{\partial}{\partial u_1^{j-1}}-\left(e^{u_2^{j-1}-u_1^{j}}+e^{u_3^{j-1}-u_2^{j}}\right)\frac{\partial}{\partial u_2^{j-1}}\\
&+e^{u_3^{j-1}-u_2^{j}}\frac{\partial}{\partial u_3^{j-1}}-\left(e^{u_2^j-u_1^{j+1}}+e^{u_2^{j+1}-u_1^{j+2}}\right)\frac{\partial}{\partial u_1^{j+2}}+\ldots,\\
K_3:=&[X_3,K]=-e^{u_3^j-u_2^{j+1}}\frac{\partial}{\partial u_2^{j+1}}+e^{u_3^j-u_2^{j+1}}\frac{\partial}{\partial u_3^{j+1}}\\
&+e^{u_3^{j-1}-u_2^{j}}\frac{\partial}{\partial u_2^{j-1}}-e^{u_3^{j-1}-u_2^{j}}\frac{\partial}{\partial u_3^{j-1}}+\ldots.
\end{align*}
Note that operators $K$ and $K_2$ are expressed in terms of operators $K_1$ and~$K_3$:
\begin{align*}
K=-K_1+K_3, \qquad K_2=-K_1-K_3.
\end{align*}
Consequently, operators $K$ and $K_2$ can be excluded from consideration. Thus, only operators $X_1$, $X_2$, $X_3$, $K_1$ and $K_3$ remain. Let us check whether this algebra is finite-dimensional. We calculate the commutators of these operators and find that only one of them yields a new operator. That is, we obtain
\begin{align*}
&[X_1,K_1]=-K_1, \qquad [X_2,K_1]=K_1, \qquad [X_3,K_1]=0,\\
&[X_1,K_3]=0, \qquad [X_2,K_3]=-K_3, \qquad [X_3,K_3]=K_3
\end{align*}
and the new operator $K_{13}$:
\begin{align*}
K_{13}:=[K_1,K_3]=&-e^{u_3^j-u_2^{j+1}+u_2^j-u_1^{j+1}}\frac{\partial}{\partial u_2^{j+1}}+e^{u_3^j-u_2^{j+1}+u_2^j-u_1^{j+1}}\frac{\partial}{\partial u_3^{j+1}}\\
&+e^{u_3^{j-1}-u_2^{j}+u_2^{j-1}-u_1^{j}}\frac{\partial}{\partial u_1^{j-1}}-e^{u_3^{j-1}-u_2^{j}+u_2^{j-1}-u_1^{j}}\frac{\partial}{\partial u_2^{j-1}}\\
&+e^{u_3^j-u_1^{j+2}}\frac{\partial}{\partial u_1^{j+2}}-e^{u_3^j-u_1^{j+2}}\frac{\partial}{\partial u_2^{j+2}}+\ldots.
\end{align*}
It follows that another operator $K_{13}$ is added to our operators $X_1$, $X_2$, $X_3$, $K_1$ and $K_3$. The characteristic algebra consisting of these six operators is finite, since their commutators do not generate new operators:
\begin{align*}
&[X_1,K_{13}]=-K_{13}, \qquad [X_2,K_{13}]=0, \qquad [X_3,K_{13}]=0,\\
&[K_1,K_{13}]=0, \qquad [K_3,K_{13}]=0.
\end{align*}
Therefore the dimension of the characteristic Lie algebra equals six.
Thus, the final system of equations for the unknown function $W$ takes the form 
\begin{align}\label{sys6W}
\begin{aligned}
&X_1W=0, \qquad K_1W=0,\\
&X_2W=0, \qquad K_3W=0,\\
&X_3W=0, \qquad K_{13}W=0.
\end{aligned}
\end{align}
Up to this point, we have assumed that the sought function $W$ depends on an infinite number of variables. To determine precisely which variables function $W$ depends on, we shall employ the scheme described in Remark 1. Using the procedure of iterating through and analyzing the variables, we find that the function $W$ depends on $u_1^j$, $u_2^j$, $u_3^j$, $u_1^{j+1}$, $u_2^{j+1}$, $u_3^{j+1}$, $u_2^{j-1}$, $u_3^{j-1}$, $u_1^{j+2}$. Then system \eqref{sys6W} turns into
\begin{align*}
&W_{u_1^j}+W_{u_1^{j+1}}+W_{u_1^{j+2}}=0,\\
&W_{u_2^j}+W_{u_2^{j+1}}+W_{u_2^{j-1}}=0,\\
&W_{u_3^j}+W_{u_3^{j+1}}+W_{u_3^{j-1}}=0,\\
&e^{u_2^j-u_1^{j+1}}W_{u_1^{j+1}}-e^{u_2^j-u_1^{j+1}}W_{u_2^{j+1}}-e^{u_2^{j-1}-u_1^{j}}W_{u_1^{j-1}}+e^{u_2^{j-1}-u_1^{j}}W_{u_2^{j-1}}\\
&\phantom{e^{u_2^j-u_1^{j+1}}W_{u_1^{j+1}}}+\left(e^{u_2^j-u_1^{j+1}}+e^{u_2^{j+1}-u_1^{j+2}}\right)W_{u_1^{j+2}}=0,\\
&e^{u_3^j-u_2^{j+1}}W_{u_2^{j+1}}-e^{u_3^j-u_2^{j+1}}W_{u_3^{j+1}}-e^{u_3^{j-1}-u_2^{j}}W_{u_2^{j-1}}+e^{u_3^{j-1}-u_2^{j}}W_{u_3^{j-1}}=0,\\
&e^{u_3^j-u_2^{j+1}+u_2^j-u_1^{j+1}}W_{u_2^{j+1}}-e^{u_3^j-u_2^{j+1}+u_2^j-u_1^{j+1}}W_{u_3^{j+1}}-e^{u_3^j-u_1^{j+2}}W_{u_1^{j+2}}\\
&\phantom{e^{u_2^j-u_1^{j+1}}W_{u_1^{j+1}}}+e^{u_3^{j-1}-u_2^{j}+u_2^{j-1}-u_1^{j}}W_{u_2^{j-1}}=0.
\end{align*}
Then by applying the consecutive integration according to the Jacobi algorithm we find that the desired function has the form
\begin{align*}
W=W(W_1,W_2,W_3),
\end{align*}
where functions $W_1$, $W_2$, $W_3$ of the form
\begin{equation}\label{W123}
\begin{aligned}
&W_1=e^{u_1^{j+1}+u_2^{j+1}+u_3^{j+1}-u_1^{j}-u_2^{j}-u_3^{j}},\\
&W_2=e^{u_2^{j}+u_3^{j}-u_2^{j-1}-u_3^{j-1}}+e^{u_1^{j+1}+u_3^{j}-u_1^{j}-u_3^{j-1}}+e^{u_1^{j+1}+u_2^{j+1}-u_1^{j}-u_2^{j}},\\
&W_3=e^{u_1^{j+2}-u_1^{j+1}}+e^{u_2^{j+1}-u_2^{j}}+e^{u_3^{j}-u_3^{j-1}}
\end{aligned}
\end{equation}
define the required complete list of $x$-integrals of the system \eqref{sysL1}. 

To use the discretization method, it is necessary to check that the integrals $W_1$-$W_3$ form a complete set of independent $x$-integrals, i.e., condition \eqref{x7} must be satisfied. Let us calculate the determinant \eqref{x7}. We obtain
\begin{equation}\label{detL1}
\begin{vmatrix}
\frac{\partial W_1}{\partial u_1^{j+1}} & \frac{\partial W_1}{\partial u_2^{j+1}} & \frac{\partial W_1}{\partial u_3^{j+1}}\\
\frac{\partial D_j(W_2)}{\partial u_1^{j+2}} & \frac{\partial D_j(W_2)}{\partial u_2^{j+2}} & \frac{\partial D_j(W_2)}{\partial u_3^{j+2}}\\
\frac{\partial D_j(W_3)}{\partial u_1^{j+3}} & \frac{\partial D_j(W_3)}{\partial u_2^{j+3}} & \frac{\partial D_j(W_3)}{\partial u_3^{j+3}}
\end{vmatrix}=
e^{u_1^{j+3}+u_2^{j+2}+u_3^{j+1}-u_1^{j}-u_2^{j}-u_3^{j}}\neq0.
\end{equation}
The determinant does not vanish, therefore functions $W_1$-$W_3$ form a complete set of independent $x$-integrals.

We now  assume that the dynamical variables depend on an additional integer $m$ such that $u_1^{j}=u_{1,m}^{j}$, $u_2^{j}=u_{2,m}^{j}$, $u_3^{j}=u_{3,m}^{j}$ and make the substitution $e^{u_{1,m}^{j}}=u_m^{j}$, $e^{u_{2,m}^{j}}=v_m^{j}$, $e^{u_{3,m}^{j}}=w_m^{j}$. Then $x$-integrals $W_1$-$W_3$ take the form
\begin{align*}
&J_1=\frac{u_m^{j+1}v_m^{j+1}w_m^{j+1}}{u_m^{j}v_m^{j}w_m^{j}},\\
&J_2=\frac{v_m^{j}w_m^{j}}{v_m^{j-1}w_m^{j-1}}+\frac{u_m^{j+1}w_m^{j}}{u_m^{j}w_m^{j-1}}+\frac{u_m^{j+1}v_m^{j+1}}{u_m^{j}v_m^{j}},\\
&J_3=\frac{u_m^{j+2}}{u_m^{j+1}}+\frac{v_m^{j+1}}{v_m^{j}}+\frac{w_m^{j}}{w_m^{j-1}}.
\end{align*}
Further we suppose that the functions $J_1$-$J_3$ are $m$-integrals of the system
\begin{equation}\label{sys-disL1}
\begin{cases}
\displaystyle u_{m+1}^{j+1}=h_1(u_{m}^{j},u_{m+1}^{j},u_{m}^{j+1},v_{m+1}^{j}), \\
\displaystyle v_{m+1}^{j+1}=h_2(v_{m}^{j},v_{m+1}^{j},v_{m}^{j+1},w_{m+1}^{j},u_{m}^{j+1}), \\
\displaystyle w_{m+1}^{j+1}=h_3(w_{m}^{j},w_{m+1}^{j},w_{m}^{j+1},v_{m}^{j+1}),
\end{cases} 
\end{equation}
which is obtained from a lattice to be determined
\begin{equation}\label{lattice-gen}
u_{n,m+1}^{j+1}=h_2(u_{n,m}^{j},u_{n,m+1}^{j},u_{n,m}^{j+1},u_{n+1,m+1}^{j},u_{n-1,m}^{j+1})
\end{equation}
by imposing special boundary conditions at the points $n=0$ and $n=4$. Below for brevity the notations are used 
\begin{equation}\label{notations}
u_{1,m}^{j}=u_m^{j},\qquad u_{2,m}^{j}=v_m^{j},\qquad u_{3,m}^{j}=w_m^{j}. 
\end{equation}
In other words, conditions $(D_m-1)J_i=0$ must be met, from which we obtain the following three equations
\begin{equation}\label{eqs}
\begin{aligned}
&(D_m-1)J_1=\frac{u_{m+1}^{j+1}v_{m+1}^{j+1}w_{m+1}^{j+1}}{u_{m+1}^{j}v_{m+1}^{j}w_{m+1}^{j}}-\frac{u_m^{j+1}v_m^{j+1}w_m^{j+1}}{u_m^{j}v_m^{j}w_m^{j}}=0,\\
&(D_m-1)J_2=\frac{v_{m+1}^{j}w_{m+1}^{j}}{v_{m+1}^{j-1}w_{m+1}^{j-1}}+\frac{u_{m+1}^{j+1}w_{m+1}^{j}}{u_{m+1}^{j}w_{m+1}^{j-1}}+\frac{u_{m+1}^{j+1}v_{m+1}^{j+1}}{u_{m+1}^{j}v_{m+1}^{j}}\\
&\phantom{(D_m-1)W_2=}-\frac{v_m^{j}w_m^{j}}{v_m^{j-1}w_m^{j-1}}-\frac{u_m^{j+1}w_m^{j}}{u_m^{j}w_m^{j-1}}-\frac{u_m^{j+1}v_m^{j+1}}{u_m^{j}v_m^{j}}=0,\\
&(D_m-1)J_3=\frac{u_{m+1}^{j+2}}{u_{m+1}^{j+1}}+\frac{v_{m+1}^{j+1}}{v_{m+1}^{j}}+\frac{w_{m+1}^{j}}{w_{m+1}^{j-1}}-\frac{u_m^{j+2}}{u_m^{j+1}}-\frac{v_m^{j+1}}{v_m^{j}}-\frac{w_m^{j}}{w_m^{j-1}}=0.
\end{aligned}
\end{equation}
By excluding in \eqref{eqs} the variables $u_{m+1}^{j+1}$, $v_{m+1}^{j+1}$, $w_{m+1}^{j+1}$, $v_{m+1}^{j-1}$, $w_{m+1}^{j-1}$ and $u_{m+1}^{j+2}$ due to \eqref{sys-disL1}
one gets  a system of equations for finding the sought functions \linebreak $h_1(u_{m}^{j},u_{m+1}^{j},u_{m}^{j+1},v_{m+1}^{j})$, $h_2(v_{m}^{j},v_{m+1}^{j},v_{m}^{j+1},w_{m+1}^{j},u_{m}^{j+1})$ and $h_3(w_{m}^{j},w_{m+1}^{j},w_{m}^{j+1},v_{m}^{j+1})$,

After the indicated transformations, the first equation of system \eqref{eqs} takes the form
\begin{equation}\label{eq1}
\begin{aligned}
&\frac{h_1(u_{m}^{j},u_{m+1}^{j},u_{m}^{j+1},v_{m+1}^{j})}{u_{m+1}^{j}}\cdot\frac{h_2(v_{m}^{j},v_{m+1}^{j},v_{m}^{j+1},w_{m+1}^{j},u_{m}^{j+1})}{v_{m+1}^{j}}\\
&\qquad \cdot\frac{h_3(w_{m}^{j},w_{m+1}^{j},w_{m}^{j+1},v_{m}^{j+1})}{w_{m+1}^{j}}-\frac{u_m^{j+1}v_m^{j+1}w_m^{j+1}}{u_m^{j}v_m^{j}w_m^{j}}=0.
\end{aligned}
\end{equation}
We will first focus on investigating this equation, so we will present the other two equations a little later. 

Differentiating equation \eqref{eq1} with respect to $u_{m+1}^{j}$ and twice with respect to $w_m^{j+1}$, we obtain the following two equations: 
\begin{align*}
&u_{m+1}^{j}\frac{\partial}{\partial u_{m+1}^{j}}h_1(u_{m}^{j},u_{m+1}^{j},u_{m}^{j+1},v_{m+1}^{j})-h_1(u_{m}^{j},u_{m+1}^{j},u_{m}^{j+1},v_{m+1}^{j})=0,\\
&\frac{\partial^2}{\partial (w_m^{j+1})^2}h_3(w_{m}^{j},w_{m+1}^{j},w_{m}^{j+1},v_{m}^{j+1})=0
\end{align*}
from which we find
\begin{align*}
&h_1(u_{m}^{j},u_{m+1}^{j},u_{m}^{j+1},v_{m+1}^{j})=h_{11}(u_{m}^{j},u_{m}^{j+1},v_{m+1}^{j})u_{m+1}^{j},\\
&h_3(w_{m}^{j},w_{m+1}^{j},w_{m}^{j+1},v_{m}^{j+1})=h_{31}(w_{m}^{j},w_{m+1}^{j},v_{m}^{j+1})w_{m}^{j+1}+h_{32}(w_{m}^{j},w_{m+1}^{j},v_{m}^{j+1}).
\end{align*}
Then, taking into account what has been found, equation \eqref{eq1} can be rewritten as
\begin{align*}
&h_{11}(u_{m}^{j},u_{m}^{j+1},v_{m+1}^{j})\cdot\frac{h_2(v_{m}^{j},v_{m+1}^{j},v_{m}^{j+1},w_{m+1}^{j},u_{m}^{j+1})}{v_{m+1}^{j}}\\
&\qquad \cdot\frac{h_{31}(w_{m}^{j},w_{m+1}^{j},v_{m}^{j+1})w_{m}^{j+1}+h_{32}(w_{m}^{j},w_{m+1}^{j},v_{m}^{j+1})}{w_{m+1}^{j}}-\frac{u_m^{j+1}v_m^{j+1}w_m^{j+1}}{u_m^{j}v_m^{j}w_m^{j}}=0.
\end{align*}
We compare the coefficients of the variable $w_{m}^{j+1}$ in the resulting relation and arrive at the following two equations: 
\begin{align}
\begin{aligned}\label{wj1}
w_{m}^{j+1}: \quad &h_{11}(u_{m}^{j},u_{m}^{j+1},v_{m+1}^{j})\cdot\frac{h_2(v_{m}^{j},v_{m+1}^{j},v_{m}^{j+1},w_{m+1}^{j},u_{m}^{j+1})}{v_{m+1}^{j}}\\
&\qquad \cdot\frac{h_{31}(w_{m}^{j},w_{m+1}^{j},v_{m}^{j+1})}{w_{m+1}^{j}}-\frac{u_m^{j+1}v_m^{j+1}}{u_m^{j}v_m^{j}w_m^{j}}=0,
\end{aligned}\\
\begin{aligned}\label{wj10}
\left(w_{m}^{j+1}\right)^0: \quad &h_{11}(u_{m}^{j},u_{m}^{j+1},v_{m+1}^{j})\cdot\frac{h_2(v_{m}^{j},v_{m+1}^{j},v_{m}^{j+1},w_{m+1}^{j},u_{m}^{j+1})}{v_{m+1}^{j}}\\
&\qquad \cdot\frac{h_{32}(w_{m}^{j},w_{m+1}^{j},v_{m}^{j+1})}{w_{m+1}^{j}}=0.
\end{aligned}
\end{align}
Since functions $h_{11}(u_{m}^{j},u_{m}^{j+1},v_{m+1}^{j})$ and $h_2(v_{m}^{j},v_{m+1}^{j},v_{m}^{j+1},w_{m+1}^{j},u_{m}^{j+1})$ are non-zero, from \eqref{wj10} we find that 
$$h_{32}(w_{m}^{j},w_{m+1}^{j},v_{m}^{j+1})=0.$$
Let us proceed to the study of equation \eqref{wj1}, which can be rewritten in one of the following forms:
\begin{align*}
&\frac{h_{11}(u_{m}^{j},u_{m}^{j+1},v_{m+1}^{j})u_m^{j}}{v_{m+1}^{j}w_{m+1}^{j}} -\frac{u_m^{j+1}v_m^{j+1}}{v_m^{j}w_m^{j}h_2(v_{m}^{j},v_{m+1}^{j},v_{m}^{j+1},w_{m+1}^{j},u_{m}^{j+1})h_{31}(w_{m}^{j},w_{m+1}^{j},v_{m}^{j+1})}=0;\\
&\frac{h_2(v_{m}^{j},v_{m+1}^{j},v_{m}^{j+1},w_{m+1}^{j},u_{m}^{j+1})v_m^{j}}{v_{m+1}^{j}w_{m+1}^{j}} -\frac{u_m^{j+1}v_m^{j+1}}{u_m^{j}w_m^{j}h_{11}(u_{m}^{j},u_{m}^{j+1},v_{m+1}^{j})h_{31}(w_{m}^{j},w_{m+1}^{j},v_{m}^{j+1})}=0;\\
&\frac{h_{31}(w_{m}^{j},w_{m+1}^{j},v_{m}^{j+1})w_m^{j}}{v_{m+1}^{j}w_{m+1}^{j}} -\frac{u_m^{j+1}v_m^{j+1}}{u_m^{j}v_m^{j}h_{11}(u_{m}^{j},u_{m}^{j+1},v_{m+1}^{j})h_2(v_{m}^{j},v_{m+1}^{j},v_{m}^{j+1},w_{m+1}^{j},u_{m}^{j+1})}=0.
\end{align*}
Next, we differentiate these equations with respect to $u_m^{j}$, $v_m^{j}$, and $w_m^{j}$, respectively, and determine the dependence of the unknown functions on these variables
\begin{align*}
&h_{11}(u_{m}^{j},u_{m}^{j+1},v_{m+1}^{j})=\frac{h_{12}(u_{m}^{j+1},v_{m+1}^{j})}{u_{m}^{j}};\\
&h_2(v_{m}^{j},v_{m+1}^{j},v_{m}^{j+1},w_{m+1}^{j},u_{m}^{j+1})=\frac{h_{22}(v_{m+1}^{j},v_{m}^{j+1},w_{m+1}^{j},u_{m}^{j+1})}{v_{m}^{j}};\\
&h_{31}(w_{m}^{j},w_{m+1}^{j},v_{m}^{j+1})=\frac{h_{32}(w_{m+1}^{j},v_{m}^{j+1})}{w_{m}^{j}}.
\end{align*}
Then, taking into account the latter, the equation \eqref{wj1} is transformed to
\begin{equation*}
\frac{h_{12}(u_{m}^{j+1},v_{m+1}^{j})h_{22}(v_{m+1}^{j},v_{m}^{j+1},w_{m+1}^{j},u_{m}^{j+1})h_{32}(w_{m+1}^{j},v_{m}^{j+1})}{v_{m+1}^{j}w_{m+1}^{j}} -u_m^{j+1}v_m^{j+1}=0.
\end{equation*}
Let us express function $h_{22}(v_{m+1}^{j},v_{m}^{j+1},w_{m+1}^{j},u_{m}^{j+1})$ from the last equality
\begin{equation*}
h_{22}(v_{m+1}^{j},v_{m}^{j+1},w_{m+1}^{j},u_{m}^{j+1}) =\frac{u_m^{j+1}v_m^{j+1}v_{m+1}^{j}w_{m+1}^{j}}{h_{12}(u_{m}^{j+1},v_{m+1}^{j})h_{32}(w_{m+1}^{j},v_{m}^{j+1})}.
\end{equation*}
Thus, using only the first equation from \eqref{eqs}, we find that the preliminary form of the sought functions is as follows
\begin{equation}\label{eq1-final}
\begin{aligned}
&h_1(u_{m}^{j},u_{m+1}^{j},u_{m}^{j+1},v_{m+1}^{j})=\frac{h_{12}(u_{m}^{j+1},v_{m+1}^{j})u_{m+1}^{j}}{u_{m}^{j}};\\
&h_2(v_{m}^{j},v_{m+1}^{j},v_{m}^{j+1},w_{m+1}^{j},u_{m}^{j+1})=\frac{u_m^{j+1}v_m^{j+1}v_{m+1}^{j}w_{m+1}^{j}}{v_{m}^{j}h_{12}(u_{m}^{j+1},v_{m+1}^{j})h_{32}(w_{m+1}^{j},v_{m}^{j+1})};\\
&h_3(w_{m}^{j},w_{m+1}^{j},w_{m}^{j+1},v_{m}^{j+1})=\frac{h_{32}(w_{m+1}^{j},v_{m}^{j+1})w_{m}^{j+1}}{w_{m}^{j}}.
\end{aligned}
\end{equation}
Now let us turn to the second equation in \eqref{eqs}, which, by virtue of \eqref{eq1-final}, takes the form
\begin{equation}\label{eq2}
\begin{aligned}
&\frac{h_{32}(w_{m+1}^{j},v_{m}^{j+1})h_{12}(u_{m}^{j+2},v_{m+1}^{j+1})}{u_{m}^{j+1}w_{m+1}^{j}}+\frac{u_{m}^{j+2}v_{m}^{j+2}h_{32}(w_{m+1}^{j},v_{m}^{j+1})}{u_{m}^{j+1}v_{m}^{j+1}h_{32}(w_{m+1}^{j+1},v_{m}^{j+2})}\\
&\qquad +\frac{u_{m}^{j+1}v_{m}^{j+1}}{v_{m}^{j}h_{12}(u_{m}^{j+1},v_{m+1}^{j})}-\frac{v_{m}^{j+1}}{v_{m}^{j}}-\frac{u_{m}^{j+2}}{u_{m}^{j+1}}-\frac{w_{m}^{j}u_{m}^{j+2}v_{m}^{j+2}}{w_{m}^{j+1}u_{m}^{j+1}v_{m}^{j+1}}=0,
\end{aligned}
\end{equation}
where
\begin{align*}
h_{12}(u_{m}^{j+2},v_{m+1}^{j+1})&=D_j(h_{12}(u_{m}^{j+1},v_{m+1}^{j}))\\
&=h_{12}\left(u_{m}^{j+2},\frac{u_m^{j+1}v_m^{j+1}v_{m+1}^{j}w_{m+1}^{j}}{v_{m}^{j}h_{12}(u_{m}^{j+1},v_{m+1}^{j})h_{32}(w_{m+1}^{j},v_{m}^{j+1})}\right),\\
h_{32}(w_{m+1}^{j+1},v_{m}^{j+2})&=D_j(h_{32}(w_{m+1}^{j},v_{m}^{j+1}))\\
&=h_{32}\left(\frac{h_{32}(w_{m+1}^{j},v_{m}^{j+1})w_{m}^{j+1}}{w_{m}^{j}},v_{m}^{j+2}\right).
\end{align*}
In equation \eqref{eq2}, only the function $h_{12}(u_{m}^{j+2},v_{m+1}^{j+1})$ contains the variable $u_{m}^{j+2}$, similarly, only $h_{32}(w_{m+1}^{j+1},v_{m}^{j+2})$ contains $v_{m}^{j+2}$. Therefore, if we differentiate equation \eqref{eq2} twice with respect to $u_{m}^{j+2}$ and, respectively, twice with respect to $v_{m}^{j+2}$, we obtain the following two equations
\begin{align*}
\frac{\partial^2}{\partial \left(u_{m}^{j+2}\right)^2}h_{12}(u_{m}^{j+2},v_{m+1}^{j+1})&=0,\\
\frac{\partial^2}{\partial \left(v_{m}^{j+2}\right)^2}h_{32}(w_{m+1}^{j+1},v_{m}^{j+2})&-\frac{2\left(\frac{\partial}{\partial v_{m}^{j+2}}h_{32}(w_{m+1}^{j+1},v_{m}^{j+2})\right)^2}{h_{32}(w_{m+1}^{j+1},v_{m}^{j+2})}\\
&+\frac{2}{v_{m}^{j+2}}\frac{\partial}{\partial v_{m}^{j+2}}h_{32}(w_{m+1}^{j+1},v_{m}^{j+2})=0.
\end{align*}
Applying the operator $D_j^{-1}$ to the last two equalities, we obtain equations for the unknown functions $h_{12}(u_{m}^{j+1},v_{m+1}^{j})$ and $h_{32}(w_{m+1}^{j},v_{m}^{j+1})$. Solving them, we find
\begin{align*}
&h_{12}(u_{m}^{j+1},v_{m+1}^{j})=h_{13}(v_{m+1}^{j})u_{m}^{j+1}+h_{14}(v_{m+1}^{j}),\\
&h_{32}(w_{m+1}^{j},v_{m}^{j+1})=-\frac{v_{m}^{j+1}}{h_{33}(w_{m+1}^{j})v_{m}^{j+1}-h_{34}(w_{m+1}^{j})}.
\end{align*}
Let us substitute the found values into equation \eqref{eq2} and collect the coefficients of the independent variables $u_{m}^{j+2}$ and $v_{m}^{j+2}$. As a result, we obtain the following three equations\begin{align*}
& 1) \quad h_{33}\Bigg(-\frac{w_{m}^{j+1}v_{m}^{j+1}}{w_{m}^{j} \left(h_{33}(w_{m+1}^{j})v_{m}^{j+1}-h_{34}(w_{m+1}^{j})\right)}\Bigg) \\
&\hphantom{2) \quad h_{13} \quad} -\frac{w_{m}^{j}\left(h_{33}(w_{m+1}^{j})v_{m}^{j+1} -h_{34}(w_{m+1}^{j})\right)}{v_{m}^{j+1}w_{m}^{j+1}}=0,
\\
& 2) \quad h_{13}\left(-\frac{u_m^{j+1}v_m^{j+1}v_{m+1}^{j}w_{m+1}^{j} \left(h_{33}(w_{m+1}^{j})v_{m}^{j+1}-h_{34}(w_{m+1}^{j})\right)}{v_{m}^{j} \left(h_{13}(v_{m+1}^{j})u_{m}^{j+1}+h_{14}(v_{m+1}^{j})\right)v_{m}^{j+1}}\right) v_{m}^{j+1}
\\
& \hphantom{2) \quad h_{13} \quad}  +h_{34}\Bigg(-\frac{w_{m}^{j+1}v_{m}^{j+1}}{w_{m}^{j} \left(h_{33}(w_{m+1}^{j})v_{m}^{j+1}-h_{34}(w_{m+1}^{j})\right)}\Bigg)w_{m+1}^{j} \\
&\hphantom{2) \quad h_{13} \quad} -\left(h_{33}(w_{m+1}^{j})v_{m}^{j+1}-h_{34}(w_{m+1}^{j})\right) w_{m+1}^{j}=0,
\\
& 3) \quad h_{14}\left(-\frac{u_m^{j+1}v_m^{j+1}v_{m+1}^{j}w_{m+1}^{j}\left(h_{33} (w_{m+1}^{j})v_{m}^{j+1}-h_{34}(w_{m+1}^{j})\right)}{v_{m}^{j} \left(h_{13}(v_{m+1}^{j})u_{m}^{j+1}+h_{14}(v_{m+1}^{j})\right)v_{m}^{j+1}}\right)
\\
&\hphantom{2) \quad h_{13} \quad} +\frac{u_{m}^{j+1}w_{m+1}^{j}\left(h_{33}(w_{m+1}^{j}) v_{m}^{j+1}-h_{34}(w_{m+1}^{j})\right)}{v_{m}^{j}}
\\
&\hphantom{2) \quad h_{13} \quad} -\frac{w_{m+1}^{j} \left(u_{m}^{j+1}\right)^2\left(h_{33}(w_{m+1}^{j})v_{m}^{j+1}-h_{34}(w_{m+1}^{j}) \right)}{v_{m}^{j}\left(h_{13}(v_{m+1}^{j})u_{m}^{j+1}+h_{14}(v_{m+1}^{j})\right)}=0.
\end{align*}
Let us analyze the second equation. Note that only the function $h_{13}$ depends on the variable $v_{m}^{j}$, and only $h_{34}$ depends on $w_{m}^{j}$. Thus, we have
\begin{align*}
h_{13}=C_1, \qquad h_{34}=C_2,
\end{align*}
where $C_1$, $C_2$ are some constants. Then the same equation allows us to determine the explicit form of the function $h_{33}$
\begin{align*}
h_{33}=-\frac{C_1}{w_{m+1}^{j}}.
\end{align*}
Substituting the found functions into the first and third equations, we find that $C_1=1$ and obtain the following equation
\begin{align*}
h_{14}\left(\frac{u_{m}^{j+1}v_{m+1}^{j}(C_2w_{m+1}^{j}+v_{m}^{j+1})}{v_{m}^{j}(u_{m}^{j+1}+h_{14}(v_{m+1}^{j}))}\right)-\frac{u_{m}^{j+1}h_{14}(v_{m+1}^{j})(C_2w_{m+1}^{j}+v_{m}^{j+1})}{v_{m}^{j}(u_{m}^{j+1}+h_{14}(v_{m+1}^{j}))}=0.
\end{align*}
Differentiating this equality twice with respect to $w_{m+1}^{j}$ and applying the operator $D_j^{-1}$ to the resulting equation, we find
\begin{align*}
h_{14}(v_{m+1}^{j})=C_3v_{m+1}^{j},
\end{align*}
where $C_3$ is an arbitrary constant. As a result we obtain that the sought functions have the following form
\begin{equation}\label{h1h2h3}
\begin{aligned}
&h_1(u_{m}^{j},u_{m+1}^{j},u_{m}^{j+1},v_{m+1}^{j})=\frac{u_{m+1}^{j}\left(u_{m}^{j+1}+C_3v_{m+1}^{j}\right)}{u_{m}^{j}},\\
&h_2(v_{m}^{j},v_{m+1}^{j},v_{m}^{j+1},w_{m+1}^{j},u_{m}^{j+1})=\frac{u_m^{j+1}v_{m+1}^{j}\left(v_{m}^{j+1}+C_2w_{m+1}^{j}\right)}{v_{m}^{j}\left(u_{m}^{j+1}+C_3v_{m+1}^{j}\right)},\\
&h_3(w_{m}^{j},w_{m+1}^{j},w_{m}^{j+1},v_{m}^{j+1})=\frac{w_{m}^{j+1}v_{m}^{j+1}w_{m+1}^{j}}{w_{m}^{j}\left(v_{m}^{j+1}+C_2w_{m+1}^{j}\right)}.
\end{aligned}
\end{equation}
Note that to determine the form of the sought functions $h_1$, $h_2$, and $h_3$, we used only two equations of the system \eqref{eqs}, therefore, it is necessary to verify whether these functions also satisfy the third equation. To do this, we substitute \eqref{h1h2h3} into the third equation of system \eqref{eqs} and verify by direct calculation that \eqref{h1h2h3} is indeed the required system satisfying all three equations \eqref{eqs}.

Thus, the sought system of equations for which functions $J_1$, $J_2$, $J_3$ are $m$-integrals has the form
\begin{equation}\label{sys-dis-L1}
\begin{cases}
\displaystyle u_{m+1}^{j+1}=\frac{u_{m+1}^{j}\left(u_{m}^{j+1}+C_3v_{m+1}^{j}\right)}{u_{m}^{j}},\\
\displaystyle v_{m+1}^{j+1}=\frac{u_m^{j+1}v_{m+1}^{j}\left(v_{m}^{j+1}+C_2w_{m+1}^{j}\right)}{v_{m}^{j}\left(u_{m}^{j+1}+C_3v_{m+1}^{j}\right)}, \\
\displaystyle w_{m+1}^{j+1}=\frac{w_{m}^{j+1}v_{m}^{j+1}w_{m+1}^{j}}{w_{m}^{j}\left(v_{m}^{j+1}+C_2w_{m+1}^{j}\right)}.
\end{cases} 
\end{equation}
Looking at the middle equation of system \eqref{sys-dis-L1}, we can conclude that the sought lattice has the form (see notations \eqref{notations} above)
\begin{equation*}
u_{n,m+1}^{j+1}=\frac{u_{n-1,m}^{j+1}u_{n,m+1}^{j}(C_2u_{n+1,m+1}^{j}+u_{n,m}^{j+1})}{u_{n,m}^{j}(C_3u_{n,m+1}^{j}+u_{n-1,m}^{j+1})},
\end{equation*}
where $C_2$ and $C_3$ are constant parameters. We specify them due to the cut-off boundary conditions $u^j_{0,m}=\infty$, $u^j_{4,m}=0$ for $m,j, \in Z$. After substitution of these conditions for $u^j_{1,m}$ and $u^j_{3,m}$ we find
\begin{align*}
u_{1,m+1}^{j+1}=\frac{u_{1,m+1}^{j}(C_2u_{2,m+1}^{j}+u_{1,m}^{j+1})}{u_{1,m}^{j}},\\
u_{3,m+1}^{j+1}=\frac{u_{2,m}^{j+1}u_{3,m+1}^{j}u_{3,m}^{j+1}}{u_{3,m}^{j}(C_3u_{3,m+1}^{j}+u_{2,m}^{j+1})}.
\end{align*}
By comparing this representations with previously found system \eqref{sys-dis-L1} and due to relations \eqref{notations} we observe that $C_3=C_2=:C$. So that 
\begin{equation*}
u_{n,m+1}^{j+1}=\frac{u_{n-1,m}^{j+1}u_{n,m+1}^{j}(Cu_{n+1,m+1}^{j}+u_{n,m}^{j+1})}{u_{n,m}^{j}(Cu_{n,m+1}^{j}+u_{n-1,m}^{j+1})}.
\end{equation*}
Let us set $C=-1$, since the constant parameter $C$ can be removed by replacing $u_{n,m}^{j}=(-C)^{-m}v_{n,m}^{j}$. Then we arrive at the well-known lattice Toda equation
\begin{equation*}
u_{n,m+1}^{j+1}=\frac{u_{n-1,m}^{j+1}u_{n,m+1}^{j}(u_{n+1,m+1}^{j}-u_{n,m}^{j+1})}{u_{n,m}^{j}(u_{n,m+1}^{j}-u_{n-1,m}^{j+1})}.
\end{equation*}
The obtained lattice is integrable, its Lax representation is given by the following system
\begin{equation*}
\begin{cases}
\displaystyle\psi^{j+1}_{n,m}=\frac{u^{j+1}_{n,m}}{u^{j}_{n,m}}\psi^{j}_{n,m}-\psi^{j}_{n+1,m},\\
\displaystyle\psi^{j}_{n,m+1}=\psi^{j}_{n,m}-\frac{u^{j}_{n,m+1}}{u^{j}_{n-1,m}}\psi^{j}_{n-1,m}.
\end{cases} 
\end{equation*}

\section{Discretization of lattice $L_2$}

Let us proceed with the lattice $L_2$ 
\begin{equation*}\label{L2}
u_{n,x}^{j+1}=u_{n,x}^{j}-e^{u_{n-1}^{j+1}-u_n^{j}}+e^{u_{n}^{j+1}-u_{n+1}^{j}}-e^{u_{n-1}^{j+1}-u_n^{j+1}}+e^{u_{n}^{j}-u_{n+1}^{j}}.
\end{equation*}
Darboux-integrable finite-field reduction of this lattice is obtained by imposing formally the cut-off boundary conditions $u^j_{0}=-\infty$, $u^j_4=\infty$ and has the form
\begin{equation*}\label{sysL2}
\begin{cases}
\displaystyle u_{1,x}^{j+1}=u^{j}_{1,x}+e^{u_1^{j+1}-v_2^j}+e^{u_1^{j}-v_2^j}, \\
\displaystyle u_{2,x}^{j+1}=v^{j}_{2,x}-e^{u_1^{j+1}-v_2^j}+e^{v_2^{j+1}-w_3^{j}}-e^{u_1^{j+1}-v_2^{j+1}}+e^{v_2^{j}-w_3^{j}}, \\
\displaystyle u_{3,x}^{j+1}=w^{j}_{3,x}-e^{v_2^{j+1}-w_3^j}-e^{v_2^{j+1}-w_3^{j+1}}.
\end{cases} 
\end{equation*}
The $x$-integrals of this system have the form
\begin{align*}
W_1=&e^{u_1^{j+1}+u_2^{j+1}+u_3^{j+1}-u_1^{j}-u_2^{j}-u_3^{j}}+e^{u_1^{j+1}+u_2^{j+1}-u_1^{j}-u_2^{j}}+e^{u_1^{j+1}+u_3^{j+1}-u_1^{j}-u_3^{j}}\\
&+e^{u_2^{j+1}+u_3^{j+1}-u_2^{j}-u_3^{j}}+e^{u_1^{j+1}-u_1^{j}}+e^{u_2^{j+1}-u_2^{j}}+e^{u_3^{j+1}-u_3^{j}},\\
W_2=&\frac{e^{u_2^{j-1}+u_3^{j-1}}}{(e^{u_2^{j}}+e^{u_2^{j-1}})(e^{u_3^{j}}+e^{u_3^{j-1}})}+\frac{e^{u_1^{j}+u_2^{j}+u_3^{j-1}}}{(e^{u_2^{j}}+e^{u_2^{j-1}})(e^{u_3^{j}}+e^{u_3^{j-1}})(e^{u_1^{j+1}}+e^{u_1^{j}})}\\
&+\frac{e^{u_1^{j}+u_2^{j}+u_3^{j}}}{(e^{u_1^{j+1}}+e^{u_1^{j}})(e^{u_2^{j+1}}+e^{u_2^{j}})(e^{u_3^{j+1}}+e^{u_3^{j}})},\\
W_3=&\frac{(e^{u_1^{j+1}}+e^{u_1^{j+2}})(e^{u_2^{j}}+e^{u_2^{j+1}})(e^{u_3^{j}}+e^{u_3^{j-1}})}{e^{u_1^{j+2}+u_2^{j+1}+u_3^{j}}}.
\end{align*}
Next, we introduce the dependence on $m$ in the following way $u_1^{j}=u_{1,m}^{j}$, $u_2^{j}=u_{2,m}^{j}$, $u_3^{j}=u_{3,m}^{j}$ and for convenience we make a replacement $e^{u_{1,m}^{j}}=u_m^{j}$, $e^{u_{2,m}^{j}}=v_m^{j}$, $e^{u_{3,m}^{j}}=w_m^{j}$. Then the $x$-integrals $W_1$-$W_3$ can be rewritten~as 
\begin{align*}
J_1=&\frac{u_m^{j+1}v_m^{j+1}w_m^{j+1}}{u_m^{j}v_m^{j}w_m^{j}}+\frac{u_m^{j+1}v_m^{j+1}}{u_m^{j}v_m^{j}}+\frac{u_m^{j+1}w_m^{j+1}}{u_m^{j}w_m^{j}}+\frac{v_m^{j+1}w_m^{j+1}}{v_m^{j}w_m^{j}}\\
&+\frac{u_m^{j+1}}{u_m^{j}}+\frac{v_m^{j+1}}{v_m^{j}}+\frac{w_m^{j+1}}{w_m^{j}},\\
J_2=&\frac{v_m^{j-1}w_m^{j-1}}{(v_m^{j}+v_m^{j-1})(w_m^{j}+w_m^{j-1})}+\frac{u_m^{j}v_m^{j}w_m^{j-1}}{(v_m^{j}+v_m^{j-1})(w_m^{j}+w_m^{j-1})(u_m^{j+1}+u_m^{j})}\\
&+\frac{u_m^{j}v_m^{j}w_m^{j}}{(u_m^{j+1}+u_m^{j})(v_m^{j+1}+v_m^{j})(w_m^{j+1}+w_m^{j})},\\
J_3=&\frac{(u_m^{j+1}+u_m^{j+2})(v_m^{j}+v_m^{j+1})(w_m^{j}+w_m^{j-1})}{u_m^{j+2}v_m^{j+1}w_m^{j}}.
\end{align*}

Let us assume that functions $J_1$-$J_3$ are also $m$-integrals of a system of the form \eqref{sys-disL1}. To find the desired system, we will use the definition of $m$-integrals, 
i.e. equalities $(D_m-1)J_i=0$, where $i=1,2,3$. Thus, from these three equalities we find the following system of equations
\begin{equation}\label{sys-dis-L2}
\begin{cases}
\displaystyle u_{m+1}^{j+1}=-\frac{u^j_{m+1}u^{j+1}_m(u^j_m-v^j_{m+1})}{u^j_m(u^{j+1}_m+v^j_{m+1})},\\
\displaystyle v_{m+1}^{j+1}=\frac{u^{j+1}_mw^j_{m+1}(v^j_m+v^{j+1}_m)-v^{j+1}_mv^j_{m+1}(v^j_m-w^j_{m+1})}{v^j_m(v^{j+1}_m+w^j_{m+1})}, \\
\displaystyle w_{m+1}^{j+1}=\frac{v^{j+1}_mw^j_m+v^{j+1}w^{j+1}_m+w^{j+1}_mw^j_{m+1}}{w^j_m}.
\end{cases} 
\end{equation}
Since system \eqref{sys-dis-L2} is a finite-field reduction of an equation of the form \eqref{lattice-gen}, the sought lattice is of the form
\begin{equation*}
u_{n,m+1}^{j+1}=\frac{u^{j+1}_{n-1,m}u^j_{n+1,m+1}(u^j_{n,m}+u^{j+1}_{n,m})-u^{j+1}_{n,m}u^j_{n,m+1}(u^j_{n,m}-u^j_{n+1,m+1})}{u^j_{n,m}(u^{j+1}_{n,m}+u^j_{n+1,m+1})}.
\end{equation*}
This lattice is known to be integrable. As a result of a linear transformation $u^j_{n,m}=v^m_{-n,j}$, it is reduced to the well-known lattice modified Kadomtsev-Petviashvili equation. The lattice modified Kadomtsev-Petviashvili equation is also a discretization of the lattice  $L_5$ (see section 9 below).

\section{Discretization of lattice $L_3$}

Let us concentrate on the lattice $L_3$
\begin{equation*}
u_{n,x}^{j+1}=u_{n,x}^{j}\frac{\left(u_n^{j+1}\right)^2}{u_{n+1}^{j}u_{n-1}^{j+1}}.
\end{equation*}
At first we impose on the lattice the special cut-off boundary conditions $u^j_{0}=c_0$, $u^j_4=c_1$. As a result we obtain the following Darboux-integrable finite-field system
\begin{equation*}\label{sysL3}
\begin{cases}
\displaystyle u_{1,x}^{j+1}=u_{1,x}^{j}\frac{\left(u_1^{j+1}\right)^2}{c_0u_2^{j}},\\
\displaystyle u_{2,x}^{j+1}=u_{2,x}^{j}\frac{\left(u_2^{j+1}\right)^2}{u_1^{j+1}u_3^{j}},\\
\displaystyle u_{3,x}^{j+1}=u_{3,x}^{j}\frac{\left(u_3^{j+1}\right)^2}{c_1u_2^{j+1}}.
\end{cases} 
\end{equation*}
The $x$-integrals of this system have the form
\begin{align*}
W_1=&\frac{u_1^j}{c_0c_1}+\frac{u_2^j}{c_1u_1^{j+1}}+\frac{u_3^j}{c_1u_2^{j+1}}+\frac{1}{u_3^{j+1}},\\
W_2=&\frac{u_3^{j-1}}{c_0c_1}+\frac{u_2^j}{c_0u_3^j}+\frac{u_1^{j+1}}{c_0u_2^{j+1}}+\frac{1}{u_1^{j+2}},\\
W_3=&\frac{u_2^{j-1}}{c_0c_1}+\frac{u_3^{j-1}u_1^j}{c_0c_1u_2^j}+\frac{u_3^{j-1}}{c_1u_1^{j+1}}+\frac{u_1^j}{c_0u_3^j}+\frac{u_2^j}{u_1^{j+1}u_3^j}+\frac{1}{u_2^{j+1}}.
\end{align*}
Assuming that variables $u_1^j$, $u_2^j$, $u_3^j$ also depend on $m$ and that functions $W_1$-$W_3$ are $m$-integrals of a system of type \eqref{sys-disL1}, we find the desired system
\begin{equation*}\label{sys-dis-L3}
\begin{cases}
\displaystyle u_{1,m+1}^{j+1}=\frac{c_0u^{j+1}_{1,m}u^j_{2,m+1}}{c_0u^j_{2,m+1}+u^j_{1,m}u^{j+1}_{1,m}-u^{j+1}_{1,m}u^j_{1,m+1}},\\
\displaystyle u_{2,m+1}^{j+1}=\frac{u^{j+1}_{1,m}u^{j+1}_{2,m}u^j_{3,m+1}}{u^{j+1}_{1,m}u^j_{3,m+1}+u^j_{2,m}u^{j+1}_{2,m}-u^{j+1}_{2,m}u^j_{2,m+1}}, \\
\displaystyle u_{3,m+1}^{j+1}=\frac{c_1u^{j+1}_{2,m}u^{j+1}_{3,m}}{c_1u^{j+1}_{2,m}+u^j_{3,m}u^{j+1}_{3,m}-u^{j+1}_{3,m}u^j_{3,m+1}},
\end{cases} 
\end{equation*}
which is a finite-field reduction of the lattice
\begin{equation*}
u_{n,m+1}^{j+1}=\frac{u^{j+1}_{n-1,m}u^{j+1}_{n,m}u^j_{n+1,m+1}}{u^{j+1}_{n-1,m}u^j_{n+1,m+1}+u^j_{n,m}u^{j+1}_{n,m}-u^{j+1}_{n,m}u^j_{n,m+1}}.
\end{equation*}
This lattice is integrable, its Lax pair is given by the following system
\begin{equation*}\label{Lax-L3}
\begin{cases}
\displaystyle\psi^{j+1}_{n,m}=\frac{u^{j+1}_{n,m}}{u^{j}_{n+1,m}}\left(\psi^{j}_{n+1,m}-\psi^{j}_{n,m}\right),\\
\displaystyle\psi^{j}_{n,m+1}=-\frac{u^{j}_{n,m+1}}{u^{j}_{n,m}}\psi^{j}_{n,m}+\left(\frac{u^{j}_{n,m+1}}{u^{j}_{n,m}}-1\right)\psi^{j}_{n-1,m}.
\end{cases} 
\end{equation*}
To the best of our knowledge the constructed lattice is new. We were unable to reduce it to any known lattice.
The Lax pair for this lattice was found by trial and error, by analogy with known Lax pairs for other equations in this class.

\section{Discretization of lattice $L_4$}

In this section we look for discretization of lattice $L_4$
\begin{equation*}
u_{n,x}^{j+1}=u_{n,x}^{j}\frac{u_{n+1}^{j}-u_n^{j+1}}{u_n^{j}-u_{n-1}^{j+1}}.
\end{equation*}
Boundary conditions $u^j_0=0$, $u^j_4=0$ reduce the lattice to the following hyperbolic system
\begin{equation*}\label{sysL4}
\begin{cases}
\displaystyle u_{1,x}^{j+1}=u_{1,x}^{j}\frac{u_2^{j}-u_1^{j+1}}{u_1^{j}},\\
\displaystyle u_{2,x}^{j+1}=u_{2,x}^{j}\frac{u_3^{j}-u_2^{j+1}}{u_2^{j}-u_1^{j+1}},\\
\displaystyle u_{3,x}^{j+1}=-u_{3,x}^{j}\frac{u_3^{j+1}}{u_3^{j}-u_2^{j+1}}.
\end{cases} 
\end{equation*}
The $x$-integrals of this Darboux-integrable system have the form
\begin{align*}
W_1=&u_1^{j-1}u_3^j(u_2^j-u_3^{j-1})(u_1^j-u_2^{j-1}),\\
W_2=&u_1^ju_1^{j+1}u_2^{j+1}-u_1^ju_2^ju_2^{j+1}+u_2^{j-1}u_2^ju_3^j-u_2^{j-1}u_3^{j-1}u_3^j,\\
W_3=&u_1^{j+1}u_1^{j+2}+u_3^{j-1}u_3^j-u_2^ju_3^j+u_2^ju_2^{j+1}-u_1^{j+1}u_2^{j+1}.
\end{align*}
We assume that the variables $u_1^j$, $u_2^j$, $u_3^j$ depend in addition on the discrete variable $m$. Then we look for a fully discrete hyperbolic system for which functions $W_1$, $W_2$, $W_3$ form a complete set of $m$-integrals. As a result we arrive at
\begin{equation*}\label{sys-dis-L4}
\begin{cases}
\displaystyle u_{1,m+1}^{j+1}=\frac{u^{j+1}_{1,m}u^j_{1,m}+u^j_{2,m+1}(u^j_{1,m+1}-u^j_{1,m})}{u^j_{1,m+1}},\\
\displaystyle u_{2,m+1}^{j+1}=\frac{u^{j+1}_{2,m}(u^j_{2,m}-u^{j+1}_{1,m})+u^j_{3,m+1}(u^j_{2,m+1}-u^j_{2,m})}{u^j_{2,m+1}-u^{j+1}_{1,m}}, \\
\displaystyle u_{3,m+1}^{j+1}=\frac{u^{j+1}_{3,m}(u^j_{3,m}-u^{j+1}_{2,m})}{u^j_{3,m+1}-u^{j+1}_{2,m}}.
\end{cases} 
\end{equation*}
It is easy to see that the resulting system is a reduction of the well-known integrable model 
\begin{equation*}
u_{n,m+1}^{j+1}=\frac{u^{j+1}_{n,m}(u^j_{n,m}-u^{j+1}_{n-1,m})+u^j_{n+1,m+1}(u^j_{n,m+1}-u^j_{n,m})}{u^j_{n,m+1}-u^{j+1}_{n-1,m}}
\end{equation*}
called the lattice Kadomtsev-Petviashvili equation. Its Lax pair is given by the following system
\begin{equation*}\label{Lax-L4}
\begin{cases}
\displaystyle\psi^{j+1}_{n,m}=\psi^{j}_{n+1,m}+(u^{j+1}_{n,m}-u^{j}_{n+1,m})\psi^{j}_{n,m},\\
\displaystyle\psi^{j}_{n,m+1}=\psi^{j}_{n,m}+\left(u^{j}_{n,m+1}-u^{j}_{n,m}\right)\psi^{j}_{n-1,m}.
\end{cases} 
\end{equation*}

\section{Discretization of lattice $L_5$}

Let us consider the lattice $L_5$
\begin{equation*}
u_{n,x}^{j+1}=u_{n,x}^{j}\frac{u_n^{j+1}\left(u_n^{j+1}-u_{n+1}^{j}\right)}{u_{n+1}^{j}\left(u_{n-1}^{j+1}-u_n^{j}\right)}.
\end{equation*}
Its Darboux-integrable reduction 
\begin{equation*}\label{sysL5}
\begin{cases}
\displaystyle u_{1,x}^{j+1}=u_{1,x}^{j}\frac{u_1^{j+1}(u_2^{j}-u_1^{j+1})}{u_1^{j}u_2^j},\\
\displaystyle u_{2,x}^{j+1}=u_{2,x}^{j}\frac{u_2^{j+1}(u_2^{j+1}-u_3^{j})}{u_3^j(u_1^{j+1}-u_2^{j})},\\
\displaystyle u_{3,x}^{j+1}=u_{3,x}^{j}\frac{u_3^{j+1}}{u_3^{j}-u_2^{j+1}}
\end{cases} 
\end{equation*}
is obtained from the lattice $L_5$ by imposing the boundary conditions $u^j_{0}=0$, $u^j_4=\infty$. Functions 
\begin{align*}
W_1=&\frac{u_1^{j-1}(u_2^j-u_3^{j-1})(u_1^j-u_2^{j-1})}{u_1^ju_2^ju_3^j},\\
W_2=&\frac{u_2^{j-1}u_3^{j-1}}{u_2^{j}u_3^{j}}-\frac{u_2^{j-1}}{u_3^{j}}+\frac{u_3^{j-1}u_1^j}{u_3^ju_1^{j+1}}-\frac{u_3^{j-1}u_1^j}{u_3^ju_2^{j}}\\
&+\frac{u_1^ju_2^j}{u_1^{j+1}u_2^{j+1}}-\frac{u_1^ju_2^j}{u_3^ju_1^{j+1}}-\frac{u_1^j}{u_2^{j+1}}+\frac{u_1^j}{u_3^j},\\
W_3=&\frac{u_3^{j-1}}{u_3^j}+\frac{u_2^j}{u_2^{j+1}}-\frac{u_2^j}{u_3^j}-\frac{u_1^{j+1}}{u_2^{j+1}}+\frac{u_1^{j+1}}{u_1^{j+2}}
\end{align*}
are $x$-integrals of this system. Functions $W_1$-$W_3$ determine the $m$-integrals of the system of equations
\begin{equation*}\label{sys-dis-L5}
\begin{cases}
\displaystyle u_{1,m+1}^{j+1}=\frac{u^{j+1}_{1,m}u^j_{1,m+1}u^j_{2,m+1}}{Cu^j_{1,m}u^{j+1}_{1,m}+u^j_{1,m}u^j_{2,m+1}+u^{j+1}_{1,m}u^j_{1,m+1}},\\
\displaystyle u_{2,m+1}^{j+1}=\frac{(Cu^{j+1}_{1,m}+u^j_{2,m+1})u^{j+1}_{2,m}u^j_{3,m+1}}{Cu^j_{2,m}u^{j+1}_{2,m}-u^{j+1}_{1,m}u^j_{3,m+1}+u^j_{2,m}u^j_{3,m+1}+u^{j+1}_{2,m}u^j_{2,m+1}}, \\
\displaystyle u_{3,m+1}^{j+1}=-\frac{(Cu^{j+1}_{2,m}+u^j_{3,m+1})u^{j+1}_{3,m}}{u^{j+1}_{2,m}-u^{j}_{3,m}},
\end{cases} 
\end{equation*}
which is a finite-field reduction of the lattice
\begin{equation*}
\bar{u}_{n,m+1}^{j+1}=\frac{(\bar{u}^{j+1}_{n-1,m}+\bar{u}^j_{n,m+1})\bar{u}^{j+1}_{n,m}\bar{u}^j_{n+1,m+1}}{\bar{u}^j_{n,m}\bar{u}^{j+1}_{n,m}-\bar{u}^{j+1}_{n-1,m}\bar{u}^j_{n+1,m+1}+\bar{u}^j_{n,m}\bar{u}^j_{n+1,m+1}+\bar{u}^{j+1}_{n,m}\bar{u}^j_{n,m+1}},
\end{equation*}
where the constant $C$ is eliminated by replacing $u^j_{n,m}=(-C)^m\bar{u}^j_{n,m}$.

By replacing $\bar{u}^j_{n,m}=\frac{1}{v^j_{n,m}}$ the found lattice is reduced to the lattice modified Kadomtsev-Petviashvili equation, which has the form:
\begin{equation*}
v_{n,m+1}^{j+1}=\frac{v^{j+1}_{n,m}v^j_{n,m+1}(v^j_{n,m}-v^{j+1}_{n-1,m})+v^j_{n+1,m+1}v^{j+1}_{n-1,m}(v^j_{n,m+1}-v^j_{n,m})}{v^j_{n,m}(v^j_{n,m+1}-v^{j+1}_{n-1,m})}.
\end{equation*}
Its Lax pair is given by the following system
\begin{equation*}\label{Lax-L5}
\begin{cases}
\displaystyle\psi^{j+1}_{n,m}=\frac{v^j_{n+1,m}}{v^{j+1}_{n,m}}(\psi^{j}_{n+1,m}+\psi^{j}_{n,m})+\psi^{j}_{n,m},\\
\displaystyle\psi^{j}_{n,m+1}=\frac{v^j_{n,m}}{v^j_{n,m+1}}(\psi^{j}_{n,m}-\psi^{j}_{n-1,m})+\psi^{j}_{n-1,m}.
\end{cases} 
\end{equation*}

\section{Discretization of lattice $L_6$}

It is easily verified that lattice $L_6$ having the form
\begin{equation*}
u^{j+1}_{n,x}=u^j_{n,x}\frac{\left(u^{j+1}_{n}-u^{j+1}_{n-1}\right)\left(u^{j+1}_{n}-u^{j}_{n+1}\right)}{\left(u^{j}_{n+1}-u^{j}_{n}\right)\left(u^{j+1}_{n-1}-u^{j}_{n}\right)}
\end{equation*}
is reduced to the system 
\begin{equation*}\label{sysL6}
\begin{cases}
\displaystyle u^{j+1}_{1,x}=u^j_{1,x}\frac{u^{j+1}_{1}\left(u^{j}_{2}-u^{j+1}_{1}\right)}{u^{j}_{1}\left(u^{j}_{2}-u^{j}_{1}\right)},\\
\displaystyle u^{j+1}_{2,x}=u^j_{2,x}\frac{\left(u^{j+1}_{2}-u^{j+1}_{1}\right)\left(u^{j+1}_{2}-u^{j}_{3}\right)}{\left(u^{j}_{3}-u^{j}_{2}\right)\left(u^{j+1}_{1}-u^{j}_{2}\right)},\\
\displaystyle u^{j+1}_{3,x}=u^j_{3,x}\frac{u^{j+1}_{3}\left(u^{j+1}_{3}-u^{j+1}_{2}\right)}{u^{j}_{3}\left(u^{j}_{3}-u^{j+1}_{2}\right)}
\end{cases} 
\end{equation*}
under the cutting off conditions $u^j_0=0$, $u^j_4=0$. The $x$-integrals of this Darboux-integrable system have the form
\begin{align*}
W_1=&\frac{(u^{j+1}_1-u^j_1)(u^{j+1}_2-u^j_2)(u^{j+1}_3-u^j_3)}{u^j_1u^{j+1}_3(u^{j+1}_1-u^{j}_2)(u^{j+1}_2-u^{j}_3)},\\
W_2=&\frac{u^{j+2}_1u^{j+1}_3\left(u^{j+1}_1(u^{j}_2-u^{j}_3)-u^j_2(u^{j+1}_2-u^{j}_3)\right)}{(u^{j+1}_1-u^{j+2}_1)(u^{j+1}_2-u^j_2)(u^{j+1}_3-u^j_3)} \\
&+\frac{u^{j+1}_1u^j_3\left(u^{j+2}_1(u^{j+2}_2-u^{j+1}_3)+u^{j+2}_2(u^{j+1}_3-u^{j+1}_2)\right)}{(u^{j+1}_1-u^{j+2}_1)(u^{j+2}_2-u^{j+1}_2)(u^{j+1}_3-u^j_3)},\\
W_3=&\frac{W_{31}}{W_{32}},
\end{align*}
where 
\begin{align*}
W_{31}=&u^{j+2}_1(u^{j+1}_2-u^{j+2}_2)(u^{j}_3-u^{j+1}_3)\left(u^{j+1}_3u^{j}_2(u^{j+2}_1-u^{j+3}_1)(u^{j+1}_2-u^{j+2}_2)(u^{j+1}_1-u^{j+1}_2)\right.\\
&\left.+u^{j+3}_1u^j_3(u^{j+1}_2-u^{j+2}_2)(u^{j+1}_1-u^{j+1}_2)(u^{j}_2-u^{j+1}_2+u^{j+1}_3)\right.\\
&\left.-u^j_3(u^{j+1}_2-u^{j+1}_3)(u^{j}_2-u^{j+1}_2)(u^{j+1}_1u^{j+2}_2+u^{j+2}_1u^{j+3}_1-u^{j+3}_1u^{j+1}_2)\right.\\
&\left.-u^{j+2}_1u^{j+1}_3u^{j}_3(u^{j+1}_2-u^{j+2}_2)(u^{j+1}_1-u^{j}_2)+u^{j+1}_1u^{j+2}_1u^{j}_3(u^{j}_2-u^{j+1}_2)(u^{j+2}_2-u^{j+1}_3)\right),\\
W_{32}=&u^{j+3}_1u^j_3(u^{j+1}_2-u^{j+1}_3)(u^{j+2}_1-u^{j+2}_2)\left(u^j_2u^{j+1}_3(u^{j+1}_2-u^{j+2}_2)(u^{j+2}_1(u^{j+1}_1-u^{j+1}_2)\right.\\
&\left.-u^j_3(u^{j+1}_1-u^{j+2}_1))+u^{j+1}_1u^j_3(u^{j+2}_1-u^{j+1}_2)(u^{j+2}_2(u^j_2-u^{j+1}_2)-u^{j+1}_3(u^j_2-u^{j+2}_2))\right).
\end{align*}

After tediously long calculations we get a hyperbolic system of the form 
\begin{equation*}\label{sys-dis-L6}
\begin{cases}
\displaystyle u_{1,m+1}^{j+1}=\frac{u^j_{1,m+1}u^{j+1}_{1,m}(u^{j}_{1,m}-u^j_{2,m+1})}{u^j_{1,m}(u^{j+1}_{1,m}-u^j_{2,m+1})+u^j_{1,m+1}(u^j_{1,m}-u^{j+1}_{1,m})},\\
\displaystyle u_{2,m+1}^{j+1}=\frac{u^j_{2,m+1}(u^j_{2,m}-u^{j+1}_{1,m})(u^{j+1}_{2,m}-u^j_{3,m+1})-u^j_{3,m+1}(u^{j+1}_{1,m}-u^j_{2,m+1})(u^j_{2,m}-u^{j+1}_{2,m})}{(u^j_{2,m}-u^{j+1}_{1,m})(u^{j+1}_{2,m}-u^j_{3,m+1})-(u^{j+1}_{1,m}-u^j_{2,m+1})(u^j_{2,m}-u^{j+1}_{2,m})},\\
\displaystyle u_{3,m+1}^{j+1}=\frac{u^j_{3,m+1}u^{j+1}_{3,m}(u^{j+1}_{2,m}-u^j_{3,m})}{u^j_{3,m}(u^{j+1}_{2,m}-u^j_{3,m+1})+u^{j+1}_{3,m}(u^j_{3,m+1}-u^{j}_{3,m})}
\end{cases} 
\end{equation*}
for which functions $W_1$-$W_3$ are $m$-integrals. 
The resulting system is a finite-field reduction of a known integrable lattice 
\begin{equation*}
u_{n,m+1}^{j+1}=\frac{u^j_{n,m+1}(u^j_{n,m}-u^{j+1}_{n-1,m})(u^{j+1}_{n,m}-u^j_{n+1,m+1})-u^j_{n+1,m+1}(u^{j+1}_{n-1,m}-u^j_{n,m+1})(u^j_{n,m}-u^{j+1}_{n,m})}{(u^j_{n,m}-u^{j+1}_{n-1,m})(u^{j+1}_{n,m}-u^j_{n+1,m+1})-(u^{j+1}_{n-1,m}-u^j_{n,m+1})(u^j_{n,m}-u^{j+1}_{n,m})}
\end{equation*}
called the Schwarzian Kadomtsev-Petviashvili equation. Its Lax pair is given by the following system
\begin{equation*}\label{Lax-L4}
\begin{cases}
\displaystyle\psi^{j+1}_{n,m}=\frac{u^{j+1}_{n,m}-u^{j}_{n+1,m}}{u^{j}_{n+1,m}-u^{j}_{n,m}}\psi^{j}_{n,m}-\frac{u^{j+1}_{n,m}-u^{j}_{n,m}}{u^{j}_{n+1,m}-u^{j}_{n,m}}\psi^{j}_{n+1,m},\\
\displaystyle\psi^{j}_{n,m+1}=\frac{u^{j}_{n,m+1}-u^{j}_{n-1,m}}{u^{j}_{n,m}-u^{j}_{n-1,m}}\psi^{j}_{n,m}-\frac{u^{j}_{n,m+1}-u^{j}_{n,m}}{u^{j}_{n,m}-u^{j}_{n-1,m}}\psi^{j}_{n-1,m}.
\end{cases} 
\end{equation*}

\section{Discretization of lattice $L_7$}

Let us consider the lattice $L_7$
\begin{equation*}
u^{j+1}_{n,x}=u^j_{n,x}+e^{u_{n-1}^{j+1}-u_n^{j+1}-u^j_n+u_{n+1}^j}
\end{equation*}
and set the boundary conditions $u_0^j=0$, $u_4^j=0$ in it, then we obtain a finite-field reduction in the form of the following Darboux-integrable system
\begin{equation*}
\begin{cases}
\displaystyle u^{j+1}_{1,x}=u^{j}_{1,x}+e^{-u_1^{j+1}-u_1^{j}+u_2^{j}}, \\
\displaystyle u^{j+1}_{2,x}=u^{j}_{2,x}+e^{-u_2^{j+1}-u_2^{j}+u_3^{j}+u_1^{j+1}}, \\
\displaystyle u^{j+1}_{3,x}=u^{j}_{3,x}+e^{-u_3^{j+1}-u_3^{j}+u_2^{j+1}}.
\end{cases} 
\end{equation*}
The $x$-integrals of this system have the form
\begin{align*}
&W_1=\frac{e^{u_1^{j+1}}e^{u_2^{j-1}}}{e^{u_1^j}e^{u_2^j}}+\frac{e^{u_1^{j-1}}}{e^{u_1^{j}}}+\frac{e^{u_2^{j+1}} e^{u_3^{j-1}}}{e^{u_2^j} e^{u_3^j}}+\frac{e^{u_3^{j+1}}}{e^{u_3^j}},\\
&W_2=\frac{e^{u_3^{j-2}}}{e^{u_3^{j-1}}}+\frac{e^{u_1^{j+2}}}{e^{u_1^{j+1}}}+\frac{e^{u_2^{j+1}}e^{u_1^{j}}}{e^{u_1^{j+1}}e^{u_2^{j}}}+\frac{e^{u_3^{j}}e^{u_2^{j-1}}}{e^{u_3^{j-1}}e^{u_2^{j}}},\\
&W_3=\frac{e^{u_2^{j}}e^{u_3^{j+1}}e^{u_1^{j+2}}}{e^{u_1^{j+1}}e^{u_2^{j+1}}e^{u_3^{j}}}+\frac{e^{u_3^{j+1}}e^{u_1^{j}}}{e^{u_1^{j+1}}e^{u_3^{j}}}+\frac{e^{u_2^{j+2}}}{e^{u_2^{j+1}}}+\frac{e^{u_3^{j-1}}e^{u_1^{j+2}}}{e^{u_1^{j+1}}e^{u_3^{j}}}+\frac{e^{u_2^{j+1}}e^{u_3^{j-1}}e^{u_1^{j}}}{e^{u_1^{j+1}}e^{u_2^{j}}e^{u_3^{j}}}+\frac{e^{u_2^{j-1}}}{e^{u_2^{j}}}.
\end{align*} 
We introduce the dependence on $m$ in the variables $u_1^{j}=u_{1,m}^{j}$, $u_2^{j}=u_{2,m}^{j}$, $u_3^{j}=u_{3,m}^{j}$ and make the replacement $e^{u_{1,m}^{j}}=v_{1,m}^{j}$, $e^{u_{2,m}^{j}}=v_{2,m}^{j}$, $e^{u_{3,m}^{j}}=v_{3,m}^{j}$. Then we assume that the functions $W_1$-$W_3$, which now have the form
\begin{align*}
&I_1=\frac{v^{j+1}_{1,m}v^{j-1}_{2,m}}{v^j_{1,m}v^j_{2,m}}+\frac{v^{j-1}_{1,m}}{v^{j}_{1,m}}+\frac{v^{j+1}_{2,m} v^{j-1}_{3,m}}{v^j_{2,m} v^j_{3,m}}+\frac{v^{j+1}_{3,m}}{v^j_{3,m}},\\
&I_2=\frac{v^{j-2}_{3,m}}{v^{j-1}_{3,m}}+\frac{v^{j+2}_{1,m}}{v^{j+1}_{1,m}}+\frac{v^{j+1}_{2,m}v^{j}_{1,m}}{v^{j+1}_{1,m}v^{j}_{2,m}}+\frac{v^{j}_{3,m}v^{j-1}_{2,m}}{v^{j-1}_{3,m}v^{j}_{2,m}},\\
&I_3=\frac{v^{j}_{2,m}v^{j+1}_{3,m}v^{j+2}_{1,m}}{v^{j+1}_{1,m}v^{j+1}_{2,m}v^{j}_{3,m}}+\frac{v^{j+1}_{3,m}v^{j}_{1,m}}{v^{j+1}_{1,m}v^{j}_{3,m}}+\frac{v^{j+2}_{2,m}}{v^{j+1}_{2,m}}+\frac{v^{j-1}_{3,m}v^{j+2}_{1,m}}{v^{j+1}_{1,m}v^{j}_{3,m}}+\frac{v^{j+1}_{2,m}v^{j-1}_{3,m}v^{j}_{1,m}}{v^{j+1}_{1,m}v^{j}_{2,m}v^{j}_{3,m}}+\frac{v^{j-1}_{2,m}}{v^{j}_{2,m}}
\end{align*} 
are $m$-integrals of the system
\begin{equation*}\label{sys-dis}
\begin{cases}
v_{1,m+1}^{j+1}=f_1(v_{1,m}^{j},v_{1,m+1}^{j},v_{1,m}^{j+1},v_{2,m+1}^{j}), \\
v_{2,m+1}^{j+1}=f_2(v_{2,m}^{j},v_{2,m+1}^{j},v_{2,m}^{j+1},v_{3,m+1}^{j},v_{1,m}^{j+1}), \\
v_{3,m+1}^{j+1}=f_3(v_{3,m}^{j},v_{3,m+1}^{j},v_{3,m}^{j+1},v_{2,m}^{j+1}).
\end{cases} 
\end{equation*}
which is obtained from the lattice
\begin{equation}\label{lattice-gen7}
v_{n,m+1}^{j+1}=f_2(v_{n,m}^{j},v_{n,m+1}^{j},v_{n,m}^{j+1},v_{n+1,m+1}^{j},v_{n-1,m}^{j+1}).
\end{equation}
From the conditions $(D_m-1)I_i=0$ we obtain the desired system of discrete equations
\begin{equation*}\label{sys-dis-L7}
\begin{cases}
v_{1,m+1}^{j+1}=\frac{v^j_{1,m+1}v^{j+1}_{1,m}+v^j_{2,m+1}}{v^j_{1,m}},\\
v_{2,m+1}^{j+1}=\frac{v^j_{2,m+1}v^{j+1}_{2,m}+v^{j+1}_{1,m}v^j_{3,m+1}}{v_{2,m}^{j}}, \\
v_{3,m+1}^{j+1}=\frac{v^j_{3,m+1}v^{j+1}_{3,m}+v^{j+1}_{2,m}}{v_{3,m}^{j}}.
\end{cases} 
\end{equation*}
Since it is assumed that the found system is obtained from a lattice of the form \eqref{lattice-gen7}, we have
\begin{equation*}
v_{n,m+1}^{j+1}=\frac{v^j_{n,m+1}v^{j+1}_{n,m}+v^{j+1}_{n-1,m}v^j_{n+1,m+1}}{v_{n,m}^{j}}.
\end{equation*}
This equation is a famous integrable lattice called the Hirota-Miwa equation. Here we present a linear system of equations associated with this lattice
\begin{equation*}\label{Lax-L7}
\begin{cases}
\displaystyle\psi^{j+1}_{n,m}=\psi^{j}_{n+1,m}+\frac{v^j_{n,m}v^{j+1}_{n+1,m}}{v^{j+1}_{n,m}v^j_{n+1,m}}\psi^{j}_{n,m},\\
\displaystyle\psi^{j}_{n,m+1}=-\psi^{j}_{n,m}+\frac{v^j_{n-1,m}v^j_{n+1,m+1}}{v^j_{n,m+1}v^j_{n,m}}\psi^{j}_{n-1,m},
\end{cases} 
\end{equation*}
The compatibility of this system is necessary for the lattice but not sufficient.

\section{Discretization of lattice $L_8$}

Let us consider the lattice $L_8$
\begin{equation*}
u_{n,x}^{j+1}=u_{n,x}^{j}+e^{u_{n+1}^{j}}+e^{u_{n-1}^{j+1}}-e^{u_{n}^{j+1}}-e^{u_{n}^{j}},
\end{equation*}
admitting Lax pair of the form
\begin{equation*}\label{Lax-L8}
\begin{cases}
\displaystyle\psi^{j}_{n,x}=\left(e^{u^j_n}-e^{u^j_{n-1}}-X^{j+1}_{n-1,x}\right)\psi^{j}_{n}-\psi^{j}_{n-1},\\
\displaystyle\psi^{j+1}_{n}=\psi^{j}_{n}+e^{u^j_n}\psi^{j}_{n+1},
\end{cases} 
\end{equation*}
where $X^j_n$ is a nonlocal variable related to function $u^j_n$ by equality $u^j_n=X^j_n-X^{j+1}_{n-1}$. This Lax pair was constructed in our recent work \cite{HabKh-25Umj} during discretization of three-dimensional lattices with one continuous and two discrete variables. 

Note that lattice  $L_8$ describes the Laplace invariants of linear semi-discrete hyperbolic equations (see \cite{AdleStartsev99}).

Discretization of lattice $L_8$ using integrals is related with certain difficulties, most likely due to the presence of a nonlocal variable in the Lax pair. In this regard, to obtain the desired lattice, we will use the available Lax pair.

Let us find a lattice of the form  
\begin{equation*}
v^{j+1}_{n,m+1}=f(v^j_{n,m},v^{j+1}_{n,m},v^j_{n,m+1},v^{j+1}_{n-1,m},v^j_{n+1,m+1})
\end{equation*}
and simultaneously determine the Lax pair 
\begin{equation}\label{Lax8pr}
\begin{cases}
\displaystyle\psi^{j+1}_{n,m}=\psi^{j}_{n,m}+e^{u^j_{n,m}}\psi^{j}_{n+1,m},\\
\displaystyle\psi^{j}_{n,m+1}=e^{A^j_{n,m}}\psi^{j}_{n,m}+B^j_{n,m}\psi^{j}_{n-1,m}.
\end{cases} 
\end{equation}
Here the sought function $v^j_{n,m}$ depends on three discrete variables $n$, $j$ and $m$. We find the unknown functional parameters $A^j_{n,m}$ and $B^j_{n,m}$ from the system compatibility condition (\ref{Lax8pr}): $D_j\psi^{j}_{n,m+1}=D_m\psi^{j+1}_{n,m}$, where $D_j$ and $D_m$ denote the operators shifting the arguments $j$ and $m$, respectively. The compatibility condition yields the following three equations: 
\begin{align*}
&1) \quad B^j_{n,m}-B^{j+1}_{n,m}=0,\\
&2) \quad e^{u^j_{n,m+1}+A^j_{n+1,m}}=e^{A^{j+1}_{n,m}+u^j_{n,m}},\\
&3) \quad e^{A^{j}_{n,m}}-e^{A^{j+1}_{n,m}}=B^{j+1}_{n,m}e^{u^j_{n-1,m}}-B^j_{n+1,m}e^{u^j_{n,m+1}}.
\end{align*}
From the first equation, we find that function $B^j_{n,m}$ does not depend on the dynamical variables. Therefore we can set $B^j_{n,m}=1$ without losing generality. 
The second equation can be rewritten as
\begin{equation*}
(D_m-1)u^j_{n,m}=(D_j-D_n)A^j_{n,m}.
\end{equation*}
In order to solve the obtained equation, we introduce a new function $h^j_{n,m}$ as follows
\begin{align*}
u^j_{n,m}=(D_j-D_n)h^j_{n,m}, \qquad A^j_{n,m}=(D_m-1)h^j_{n,m}.
\end{align*}
Then in the new variables $h^j_{n,m}$ the third equation can be rewritten as
\begin{align*}
e^{h^j_{n,m+1}-h^j_{n,m}}-e^{h^{j+1}_{n,m+1}-h^{j+1}_{n,m}}=e^{h^{j+1}_{n-1,m}-h^j_{n,m}}-e^{h^{j+1}_{n,m+1}-h^j_{n+1,m+1}}.
\end{align*}
Let us make a point transformation of the variables $e^{h^j_{n,m}}=H^j_{n,m}$ in the last equation, then we obtain the lattice  
\begin{equation}\label{wToda}
H^{j+1}_{n,m+1}=\frac{H_{n+1,m+1}^{j}H_{n,m}^{j+1}(H_{n-1,m}^{j+1}-H_{n,m+1}^{j})}{H_{n,m}^{j}(H_{n,m}^{j+1}-H_{n+1,m+1}^{j})}.
\end{equation}
By rotating the coordinate axes of the independent variables $H_{n,m}^{j}=\bar{H}_{-n,j}^{m}$, the last equation will be transformed into the standard form of lattice Toda equation. We obtain the Lax pair of this lattice by substituting the found parameter values $u^j_{n,m}$, $A^j_{n,m}$ and $B^j_{n,m}$ into system \eqref{Lax8pr}:
\begin{equation*}
\begin{cases}
\displaystyle\psi^{j+1}_{n,m}=\psi^{j}_{n,m}+\frac{H^{j+1}_{n,m}}{H^j_{n+1,m}}\psi^{j}_{n+1,m},\\
\displaystyle\psi^{j}_{n,m+1}=\frac{H^{j}_{n,m+1}}{H^j_{n,m}}\psi^{j}_{n,m}+\psi^{j}_{n-1,m}.
\end{cases} 
\end{equation*}
Using the substitution $v^j_{n,m+1}=\frac{H^{j+1}_{n-1,m}}{H^j_{n,m+1}}$ the lattice \eqref{wToda} is reduced to the form
\begin{equation}\label{obtained}
v^{j+1}_{n,m+1}=\frac{v^{j}_{n,m+1}v^{j+1}_{n,m}(v^{j+1}_{n-1,m}-1)(v^{j}_{n+1,m+1}-1)}{v^{j}_{n,m}(v^{j}_{n,m+1}-1)(v^{j+1}_{n,m}-1)}.
\end{equation}

Let us evaluate the continuum limit of the obtained lattice \eqref{obtained}. To this end we assume that $x=m\varepsilon$, $v^{j}_{n,m}=\varepsilon s^{j}_{n,m}$. Thus we have 
\begin{equation}\label{ttt}
s^{j}_{n,m+1}=s^{j}_{n,m}+\varepsilon s^{j}_{n,x} +O(\varepsilon^2), \quad \mbox{for} \quad \varepsilon \rightarrow 0.
\end{equation}
Now the lattice (\ref{obtained}) takes the form

\begin{equation}\label{changed}
     \frac{s^{j+1}_{n,m+1}}{s^{j+1}_{n,m}}=\frac{s^{j}_{n,m+1}} {s^{j}_{n,m}} \frac{ (1-\varepsilon s^{j+1}_{n-1,m})(1-\varepsilon s^{j}_{n+1,m+1})}{(1-\varepsilon s^{j}_{n,m+1})(1- \varepsilon s^{j+1}_{n,m}-1)}.
\end{equation}
Then we get rid the variables with the shifted values of $m$:  $s^{j+1}_{n,m+1}$, $s^{j}_{n,m+1} $, $s^{j}_{n+1,m+1} $, $s^{j}_{n,m+1}$ due to relation \eqref{ttt}. Afterwards we take the logarithm of the both sides of \eqref{changed} and  simplify due to the formula $\ln (1-\varepsilon s)=-\varepsilon s+ O(\varepsilon^2)$ for $\varepsilon \rightarrow 0$. As a result we get 
$$\frac{s^{j+1}_{n,x}}{s^{j+1}_{n,m}}=\frac{s^{j}_{n,x}} {s^{j}_{n,m}} -s^{j+1}_{n-1,m}-s^{j}_{n+1,m} +s^{j}_{n,m}+s^{j+1}_{n,m}+ O(\varepsilon),$$
for $\varepsilon \rightarrow 0$. Then we set  $\varepsilon=0$ and by replacing $s^{j}_{n,m}=\exp {u^j_n}$ arrive at the lattice $L_8$.

The found above lattice (\ref{obtained}), by using transformations, is reduced to a well-known lattice called the Y-system. Let us discuss this point in more detail. First, we make a linear substitution of the form $v^{j}_{n,m}=\bar{v}^{j}_{-n,m}+1$ and then we obtain
\begin{equation*}
\bar{v}^{j+1}_{-n,m+1}=\frac{(\bar{v}^{j}_{-n,m+1}+1)(\bar{v}^{j+1}_{-n,m}+1)\bar{v}^{j+1}_{-n+1,m}\bar{v}^{j}_{-n-1,m+1}}{(\bar{v}^{j}_{-n,m}+1)\bar{v}^{j}_{-n,m+1}\bar{v}^{j+1}_{-n,m}}-1.
\end{equation*}
Next, using the linear substitution $\bar{v}^{j}_{-\bar{n},m} = w^{j}_{\hat{n}-j+m,m}$, we rewrite this lattice as follows
\begin{equation*}
w^{j+1}_{\hat{n}-j+m,m+1}=\frac{(w^{j}_{\hat{n}-j+m+1,m+1}+1)(w^{j+1}_{\hat{n}-j-1+m,m}+1)w^{j+1}_{\hat{n}-j+m,m}w^{j}_{\hat{n}-j+m,m+1}}{(w^{j}_{\hat{n}-j+m,m}+1)w^{j}_{\hat{n}-j+m+1,m+1}w^{j+1}_{\hat{n}-j-1+m,m}}-1.
\end{equation*}
Now we make the final change of variables $\tilde{n}=\hat{n}-j+m$ and obtain the desired Y-system:
\begin{equation}\label{Y-sys}
w^{j+1}_{\tilde{n},m+1}=\frac{(w^{j}_{\tilde{n}+1,m+1}+1)(w^{j+1}_{\tilde{n}-1,m}+1)w^{j+1}_{\tilde{n},m}w^{j}_{\tilde{n},m+1}}{(w^{j}_{\tilde{n},m}+1)w^{j}_{\tilde{n}+1,m+1}w^{j+1}_{\tilde{n}-1,m}}-1.
\end{equation}
For the lattice \eqref{Y-sys} we present an associated system of equations of the form
\begin{equation*}
\begin{cases}
\displaystyle\varphi^{j+1}_{n,m}=\left(\frac{w^j_{n+1,m+1}+1}{w^j_{n,m}+1}+\frac{w^j_{n+1,m+1}}{w^{j+1}_{n,m}}\right)\varphi^{j}_{n+1,m}+\varphi^{j}_{n,m},\\
\displaystyle\varphi^{j}_{n,m+1}=\varphi^{j}_{n-1,m}+\frac{w^{j}_{n,m+1}+1}{w^j_{n-1,m}+1}\varphi^{j}_{n,m}.
\end{cases} 
\end{equation*}
The compatibility condition of the system is sufficient for \eqref{Y-sys}, but not necessary.


\section{Discretization of lattice $L_9$}
In this section we search the discretization of lattice
\begin{equation*}
u_{n,x}^{j+1}=u_{n,x}^{j}+\left(u_n^{j+1}-u_n^{j}\right)\left(u_n^{j+1}+u_n^{j}-u_{n-1}^{j+1}-u_{n+1}^{j}\right)
\end{equation*}
using its Lax pair
\begin{equation*}\label{Lax-L9}
\begin{cases}
\displaystyle \psi^{j}_{n,x}=(u^j_{n-1}-u^j_n)\psi^{j}_{n}+\psi^{j}_{n-1},\\
\displaystyle \psi^{j+1}_{n}=(u^j_n-u^{j+1}_n)\psi^{j}_{n+1}-\psi^{j}_{n}
\end{cases} 
\end{equation*}
since the integrals of the reduced system are very huge and complicated. 
We will look for a lattice of the form
\begin{equation*}
u^{j+1}_{n,m+1}=f(u^j_{n,m},u^{j+1}_{n,m},u^j_{n,m+1},u^{j+1}_{n-1,m},u^j_{n+1,m+1})
\end{equation*}
for which the Lax pair is defined by the following system of compatible equations
\begin{equation*}
\begin{cases}
\displaystyle\psi^{j+1}_{n,m}=(u^j_{n,m}-u^{j+1}_{n,m})\psi^{j}_{n+1,m}-\psi^{j}_{n,m},\\
\displaystyle\psi^{j}_{n,m+1}=A(j,n,m)\psi^{j}_{n,m}+B(j,n,m)\psi^{j}_{n-1,m},
\end{cases} 
\end{equation*}
where $A(j,n,m)$, $B(j,n,m)$ are unknown functions. From the compatibility condition of the Lax equations
\begin{align*}
&D_m\psi^{j+1}_{n,m}-D_j\psi^{j}_{n,m+1}=\\
&D_m\left((u^j_{n,m}-u^{j+1}_{n,m})\psi^{j}_{n+1,m}-\psi^{j}_{n,m}\right)-D_j\left(A(j,n,m)\psi^{j}_{n,m}+B(j,n,m)\psi^{j}_{n-1,m}\right)=0
\end{align*}
we find the unknown functions
\begin{align*}
&A(j,n,m)=u^j_{n,m+1}-u^j_{n-1,m}, \qquad B(j,n,m)=1,\\
&f=\frac{u^j_{n,m+1}(u^j_{n,m}-u^j_{n+1,m+1})+u^{j+1}_{n-1,m}(u^{j+1}_{n,m}-u^j_{n,m})}{u^{j+1}_{n,m}-u^j_{n+1,m+1}}.
\end{align*}
Thus, the desired discretization of the lattice $L_9$ has the form
\begin{align*}
u^{j+1}_{n,m+1}=\frac{u^j_{n,m+1}(u^j_{n,m}-u^j_{n+1,m+1})+u^{j+1}_{n-1,m}(u^{j+1}_{n,m}-u^j_{n,m})}{u^{j+1}_{n,m}-u^j_{n+1,m+1}}
\end{align*}
and its Lax pair is given by the following system of equations
\begin{equation*}\label{Lax-L9dis}
\begin{cases}
\displaystyle\psi^{j+1}_{n,m}=(u^j_{n,m}-u^{j+1}_{n,m})\psi^{j}_{n+1,m}-\psi^{j}_{n,m},\\
\displaystyle\psi^{j}_{n,m+1}=(u^j_{n,m+1}-u^j_{n-1,m})\psi^{j}_{n,m}+\psi^{j}_{n-1,m}.
\end{cases} 
\end{equation*}
The found lattice is known to be integrable. By the linear substitution $u^j_{n,m}=v^m_{-n,j}$ it can be reduced to the standard form of the lattice Kadomtsev-Petviashvili equation (this lattice is a discretization of the lattice $L_4$ as well (see~\S 8)).

\section{List of discrete models of Hirota-Miwa type}

During the discretization of lattices of the form \eqref{hirotatype}, the following list of equations was obtained:
\begin{enumerate}
	\item[$HM_1$)] $u_{n,m+1}^{j+1}=\frac{u_{n-1,m}^{j+1}u_{n,m+1}^{j}(u_{n+1,m+1}^{j}-u_{n,m}^{j+1})}{u_{n,m}^{j}(u_{n,m+1}^{j}-u_{n-1,m}^{j+1})}$ (lattice Toda equation);
	\item[$HM_2$)] $u^{j+1}_{n,m+1}=\frac{(u^{j}_{n+1,m+1}+1)(u^{j+1}_{n-1,m}+1)u^{j+1}_{n,m}u^{j}_{n,m+1}}{(u^{j}_{n,m}+1)u^{j}_{n+1,m+1}u^{j+1}_{n-1,m}}-1$ (Y-system);
	\item[$HM_3$)] $u_{n,m+1}^{j+1}=\frac{u^{j+1}_{n,m}(u^j_{n,m}-u^{j+1}_{n-1,m})+u^j_{n+1,m+1}(u^j_{n,m+1}-u^j_{n,m})}{u^j_{n,m+1}-u^{j+1}_{n-1,m}}$ (lattice Kadomtsev-Petviashvili equation);
	\item[$HM_4$)] $u_ {n,m+1}^{j+1}=\frac{u^{j+1}_{n,m}u^j_{n,m+1}(u^j_{n,m}-u^{j+1}_{n-1,m})+u^j_{n+1,m+1}u^{j+1}_{n-1,m}(u^j_{n,m+1}-u^j_{n,m})}{u^j_{n,m}(u^j_{n,m+1}-u^{j+1}_{n-1,m})}$ (lattice modified Kadomtsev-Petviashvili equation);
	\item[$HM_5$)] $u_{n,m+1}^{j+1}=\frac{u^j_{n,m+1}(u^j_{n,m}-u^{j+1}_{n-1,m})(u^{j+1}_{n,m}-u^j_{n+1,m+1})-u^j_{n+1,m+1}(u^{j+1}_{n-1,m}-u^j_{n,m+1})(u^j_{n,m}-u^{j+1}_{n,m})}{(u^j_{n,m}-u^{j+1}_{n-1,m})(u^{j+1}_{n,m}-u^j_{n+1,m+1})-(u^{j+1}_{n-1,m}-u^j_{n,m+1})(u^j_{n,m}-u^{j+1}_{n,m})}$ (Schwarzian Kadomtsev-Petviashvili equation);
	\item[$HM_6$)] $u_{n,m+1}^{j+1}=\frac{u^j_{n,m+1}u^{j+1}_{n,m}+u^{j+1}_{n-1,m}u^j_{n+1,m+1}}{u_{n,m}^{j}}$ (Hirota-Miwa equation);
	\item[$HM_7$)] $u_{n,m+1}^{j+1}=\frac{u^{j+1}_{n-1,m}u^{j+1}_{n,m}u^j_{n+1,m+1}}{u^{j+1}_{n-1,m}u^j_{n+1,m+1}+u^j_{n,m}u^{j+1}_{n,m}-u^{j+1}_{n,m}u^j_{n,m+1}}$ (presumably a new lattice).
\end{enumerate}

Let us briefly describe the correspondence between the known equations of the classes \eqref{sdtodatype} and \eqref{hirotatype}. The lattice $HM_1$ is a discretization of the equation $L_1$, the lattice $HM_2$ is a discretization of the equation $L_8$, the lattice $HM_3$ is a discretization of the equation $L_4$ and simultaneously discretization of $L_9$, the lattice $HM_4$ is a discretization of the equation $L_2$ and simultaneously discretization of $L_5$, the lattice $HM_5$ is a discretization of the equation $L_6$, the lattice $HM_6$ is a discretization of the equation $L_7$, the lattice $HM_7$ is a discretization of the equation $L_3$.

For lattices $HM_1$, $HM_3$-$HM_5$ and $HM_7$ Lax pairs are constructed, and for lattices $HM_6$ and $HM_2$, associated systems (the compatibility of the system (6) is necessary for the lattice $HM_6$ but not sufficient, the compatibility condition of the system (7) is sufficient for $HM_2$, but not necessary):
\begin{align*}
&(1) \qquad \begin{cases} 
\displaystyle\psi^{j+1}_{n,m}=\frac{u^{j+1}_{n,m}}{u^{j}_{n,m}}\psi^{j}_{n,m}-\psi^{j}_{n+1,m},\\
\displaystyle\psi^{j}_{n,m+1}=\psi^{j}_{n,m}-\frac{u^{j}_{n,m+1}}{u^{j}_{n-1,m}}\psi^{j}_{n-1,m};
\end{cases} \\
&(2) \qquad \begin{cases}
\displaystyle\psi^{j+1}_{n,m}=\left(\frac{u^j_{n+1,m+1}+1}{u^j_{n,m}+1}+\frac{u^j_{n+1,m+1}}{u^{j+1}_{n,m}}\right)\psi^{j}_{n+1,m}+\psi^{j}_{n,m},\\
\displaystyle\psi^{j}_{n,m+1}=\psi^{j}_{n-1,m}+\frac{u^{j}_{n,m+1}+1}{u^j_{n-1,m}+1}\psi^{j}_{n,m};
\end{cases}  \\
&(3) \qquad \begin{cases} 
\displaystyle\psi^{j+1}_{n,m}=\psi^{j}_{n+1,m}+(u^{j+1}_{n,m}-u^{j}_{n+1,m})\psi^{j}_{n,m},\\
\displaystyle\psi^{j}_{n,m+1}=\psi^{j}_{n,m}+\left(u^{j}_{n,m+1}-u^{j}_{n,m}\right)\psi^{j}_{n-1,m};
\end{cases} \\
&(4) \qquad \begin{cases} 
\displaystyle\psi^{j+1}_{n,m}=\frac{u^j_{n+1,m}}{u^{j+1}_{n,m}}(\psi^{j}_{n+1,m}+\psi^{j}_{n,m})+\psi^{j}_{n,m},\\
\displaystyle\psi^{j}_{n,m+1}=\frac{u^j_{n,m}}{u^j_{n,m+1}}(\psi^{j}_{n,m}-\psi^{j}_{n-1,m})+\psi^{j}_{n-1,m};
\end{cases} \\
&(5) \qquad \begin{cases}
\displaystyle\psi^{j+1}_{n,m}=\frac{u^{j+1}_{n,m}-u^{j}_{n+1,m}}{u^{j}_{n+1,m}-u^{j}_{n,m}}\psi^{j}_{n,m}-\frac{u^{j+1}_{n,m}-u^{j}_{n,m}}{u^{j}_{n+1,m}-u^{j}_{n,m}}\psi^{j}_{n+1,m},\\
\displaystyle\psi^{j}_{n,m+1}=\frac{u^{j}_{n,m+1}-u^{j}_{n-1,m}}{u^{j}_{n,m}-u^{j}_{n-1,m}}\psi^{j}_{n,m}-\frac{u^{j}_{n,m+1}-u^{j}_{n,m}}{u^{j}_{n,m}-u^{j}_{n-1,m}}\psi^{j}_{n-1,m};
\end{cases}  \\
&(6) \qquad \begin{cases}
\displaystyle\psi^{j+1}_{n,m}=\psi^{j}_{n+1,m}+\frac{u^j_{n,m}u^{j+1}_{n+1,m}}{u^{j+1}_{n,m}u^j_{n+1,m}}\psi^{j}_{n,m},\\
\displaystyle\psi^{j}_{n,m+1}=-\psi^{j}_{n,m}+\frac{u^j_{n-1,m}u^j_{n+1,m+1}}{u^j_{n,m+1}u^j_{n,m}}\psi^{j}_{n-1,m};
\end{cases} \\
&(7) \qquad \begin{cases} 
\displaystyle\psi^{j+1}_{n,m}=\frac{u^{j+1}_{n,m}}{u^{j}_{n+1,m}}\left(\psi^{j}_{n+1,m}-\psi^{j}_{n,m}\right),\\
\displaystyle\psi^{j}_{n,m+1}=-\frac{u^{j}_{n,m+1}}{u^{j}_{n,m}}\psi^{j}_{n,m}+\left(\frac{u^{j}_{n,m+1}}{u^{j}_{n,m}}-1\right)\psi^{j}_{n-1,m}.
\end{cases}
\end{align*}

By discretizing the semi-discrete equations $L_1$-$L_9$, we obtained a list of seven Hirota-Miwa type equations, unrelated by point transformations of the variables. Six of these equations (see $HM_1$-$HM_6$) are contained in previously compiled lists (see \cite{Hirota}-\cite{Ferapontov}). We were unable to reduce one of the equations (see $HM_7$) to known equations; it may be new.

It should be noted that in addition to these six, seven more integrable equations of the Hirota-Miwa type are considered in the literature.
They were obtained by different authors at different times, starting in the 1980s (see \cite{Hirota}-\cite{Ferapontov}). However, it turned out that these seven equations can be reduced by point transformations of the variables to the six equations in the list mentioned above (see $HM_1$-$HM_6$). We observed that <<Lattice spin equation>>, <<Schwarzian Toda equation>> and <<Sine-Gordon equation>> reduce to $HM_5$, <<Toda equation for rotation coefficients>> and <<One more version of the Toda equation>> to $HM_1$, <<Lattice modified Toda equation>> to $HM_4$, and <<Toda equation>> to $HM_6$.

\section*{Conclusions}

This paper convincingly demonstrates that integrable discrete nonlinear equations with three independent variables of all three types (\ref{todatype}), (\ref{sdtodatype}) and (\ref{hirotatype}) possess the following characteristic property. For each of these models, there exists a unique termination condition which, when imposed at two points on a distinguished integer line, reduces the lattice to a finite hyperbolic system with enhanced integrability. More precisely, these systems admit complete sets of characteristic integrals in both directions, i.e., they are integrable in the Darboux sense.
It is noteworthy that there is a close connection between classes (\ref{todatype}), (\ref{sdtodatype}) and (\ref{hirotatype}), which is established at the level of finite-field reductions. In the recent works \cite{GarHab-25PhysD} and \cite{HabKh-26RChD} the finite-field reduction of lattices $E_6$ and $E_3$ (see the list above) were used to construct new explicit solutions of the Ishimori and respectively, Davey-Stewartson equations. 
A classification algorithm using the well-known symmetry approach in combination with the finite-field reduction method is proposed in \cite{GarHab-25JPA}. The algorithm can be applied to the problem of complete classification  of integrable lattices of the form (\ref{todatype})-(\ref{hirotatype}).

\section*{Acknowledgments}

This work is supported by the Russian Science Foundation, grant no. 25-21-00050, https://rscf.ru/project/25-21-00050/.

\section*{CRediT authorship contribution statement}

Ismagil T. Habibullin: Writing – review \& editing, Writing – original draft, Validation, Methodology, Investigation, Formal analysis. Aigul R. Khakimova: Writing – review \& editing, Writing – original draft, Validation, Investigation, Formal analysis.




{\color{red}[ changed]}

\end{document}